\documentclass[12pt]{article}

\usepackage{newtxtext,newtxmath}

\usepackage{graphicx}
\usepackage{caption}
\usepackage[letterpaper,margin=1in]{geometry}
\usepackage[section]{placeins} 
\usepackage{afterpage}
\let\OldFloatBarrier\FloatBarrier
\renewcommand{\FloatBarrier}{\let\OldFloatBarrier\FloatBarrier} 
\renewenvironment{abstract}
	{\quotation}
	{\endquotation}

\date{}

\makeatletter
\renewcommand{\fnum@figure}{\textbf{Figure \thefigure}}
\renewcommand{\fnum@table}{\textbf{Table \thetable}}
\makeatother

\usepackage{scicite}

\usepackage{url}
\usepackage[colorlinks=false, pdfborder={0 0 0}]{hyperref}
\hypersetup{breaklinks=true}

\def\scititle{Density and sea level changes weaken the Atlantic Meridional Overturning Circulation through ocean bottom pressure}
\title{\bfseries \boldmath \scititle}

\author{
	Qianjiang~Xing$^{1}$,
	Shane~Elipot$^{1\ast}$,
	William~E. Johns$^{1}$,
    Bablu Sinha$^{2}$,\and 
    Jules Kajtar$^{2}$,
    Adam~T. Blaker$^{2}$,
    Tillys Petit$^{2}$,
    Ben~I. Moat$^{2}$, and
    David~A. Smeed$^{2}$
    \\
	\small$^{1}$Rosenstiel School of Marine, Atmospheric, and Earth Science, University of Miami, Miami, USA.\and
	\small$^{2}$National Oceanography Centre, Southampton, United Kingdom.\and
	\small$^\ast$Corresponding author. Email: selipot@miami.edu\and
}

\begin{document} 

\maketitle

\begin{abstract} \bfseries \boldmath \small
Observations from the RAPID array at 26.5$^{\circ}$N indicate a linear decline in the Atlantic Meridional Overturning Circulation (AMOC) over the past two decades, linked to contrasting boundary changes: a weakening western-boundary contribution that is partly compensated by strengthening at the eastern boundary. Yet it remains unclear whether this partial compensation reflects a basin-wide adjustment or a regional feature, and what processes drive it. Here we use a high-resolution ocean model to investigate the spatial structure and underlying mechanisms of the AMOC change across the mid-latitude North Atlantic. The model reproduces a meridionally coherent decline in western-boundary deep overturning transport together with a partially compensating strengthening at the eastern boundary, consistent with observations at 26.5$^{\circ}$N. These opposing trends arise from a vertically coherent ocean bottom pressure trend shaped by two competing drivers: rising coastal sea level and decreasing interior density. Through geostrophic balance, this mechanism produces partial boundary compensation across latitudes, yielding a basin-wide AMOC decline throughout the mid-latitude North Atlantic.
\end{abstract}

\section*{Main}

In the North Atlantic, the Atlantic Meridional Overturning Circulation (AMOC) consists of an upper limb that carries warm surface and thermocline waters northward and a lower limb that returns cold, dense deep waters southward. The strength of this vertically sheared circulation is commonly quantified by the overturning streamfunction, which represents the cumulative meridional transport above and below the depth of maximum overturning. In other word, the AMOC strength can be approximated by the overturning transport integrated either from the surface to the depth of the overturning streamfunction maximum (hereafter, the AMOC depth) or from the AMOC depth to the depth at which the overturning streamfunction crosses zero (hereafter, the zero-crossing overturning depth).\\

Directly monitoring the AMOC remains challenging because the overturning circulation spans the entire Atlantic basin and is embedded within energetic mesoscale variability that dominates local current measurements \cite{ferrari2009ocean,hughes2018window}. Because the AMOC is fundamentally a basin-scale geostrophic circulation, changes in overturning strength are associated with coherent pressure anomalies along the continental boundaries. This relationship has motivated the use of ocean bottom pressure (OBP) gradients to estimate overturning transports \cite{bingham2008determining,hughes2013test,elipot2013coherence,elipot2014observed}. Under geostrophic balance, vertically integrated meridional transport can be inferred from pressure gradient differences between the eastern and western basin boundaries. When applied below the depth of maximum overturning, the resulting deep overturning transport provides a close approximation to AMOC variability. Modeling and observational studies further show that boundary OBP measurements effectively filter mesoscale variability and capture the large-scale, climate-relevant overturning signal \cite{bingham2008determining,elipot2014observed,hughes2018window}. Because OBP variations are coherent across much of the Atlantic basin \cite{bryden2009adjustment,elipot2013coherence}, they offer a powerful framework for investigating basin-scale overturning connectivity and variability.\\

A previous observational study generated four short-term (3.6-year) time series of the deep western overturning transport relative to and below 1000 m, calculated through double vertical integration of western boundary OBP cross-slope (normal to bathymetric contours) gradients or equivalent geostrophic transport shear, based on four mooring arrays deployed at different latitudes across the North Atlantic \cite{elipot2017observed}. Their analysis revealed that short-term variations in deep western overturning transport were meridionally coherent and largely driven by wind stress \cite{elipot2017observed}. Yet, the brevity of these records prevented any robust assessment of long-term trends or decadal variability at the western boundary. A recent multi-latitude study revisited these four time series over a much longer period, up to 22 years, and revealed a meridionally consistent decline in deep western overturning transport over the past two decades \cite{xing2026meridionally}. This meridionally consistent signal provides strong evidence of a basin-scale weakening of the AMOC. Moreover, using RAPID array observations at 26.5$^{\circ}$N, that study identified an opposing-trend pattern between western and eastern boundary contributions to the AMOC. This is an important result that prompts the question of what mechanisms drive this contrasting boundary trend pattern under current climatic conditions. In addition, it is reasonable to ask whether the opposing-trend pattern observed by the RAPID array is representative of the broader North Atlantic.\\

To address the question regarding the consistency of opposing trends, obtaining abundant OBP data across the wider North Atlantic is critical. However, such data are currently unavailable as the existing OBP observations suffer from severe limitations. Bottom pressure recorders, although they directly measure OBP and excel at capturing high-frequency variability, suffer from long-period drift which precludes their use for the assessment of climate trends \cite{hughes2013test}. As a result, in-situ OBP observations offer limited mechanistic insight into the opposing boundary trend pattern. In addition, few observational arrays provide basin-wide geostrophic transport profiles needed to estimate overturning transport on both the western and eastern boundaries \cite{mccarthy2020sustainable}, with the RAPID array being an exception. Furthermore, the temporal coverage across the existing arrays at the western boundary varies in record length. For example, although the four time series of deep western overturning transport, estimated by recent observational study \cite{xing2026meridionally}, display a coherent meridional decline, only the two southern arrays of the four mooring arrays, RAPID and MOVE at 26.5$^{\circ}$N and 16.5$^{\circ}$N, respectively, have been operating for nearly two decades \cite{moat2024atlantic,send2011observation}. In contrast, the two northern arrays, Line W at 39.5$^{\circ}$N and the RAPID–Scotian Line at 42.5$^{\circ}$N, provide only about ten years of continuous observations \cite{toole2017moored,LineRS}. Consequently, our ability using observational data to fully characterize the meridional consistency of the opposing trends across the broader North Atlantic remains limited.\\

To overcome these observational limitations, we utilize 24 years (2000-2023) of simulated data from a realistically forced high-resolution ocean model to investigate the trends of deep overturning transports inferred from OBP cross-slope gradients (see Materials and Methods) across the mid-latitude North Atlantic (25–41$^{\circ}$N). First, we compute linear trends of deep overturning transports at both western and eastern boundaries to test the robustness and meridional consistency of the observed opposing-trend pattern, as well as reassess their relationship with AMOC derived from the velocity field. Second, we explore the physical mechanisms underlying this opposing boundary trend by examining the spatial structure of the simulated OBP trends. Finally, we identify the processes driving the spatial pattern of the OBP trend under contemporary climate change conditions as represented in the model (model validation in terms of climate change is provided in Materials and Methods).\\

To maintain consistency with our previous study \cite{xing2026meridionally}, deep overturning transports and their western and eastern contributions presented in this study are calculated by vertically integrating OBP gradients between the AMOC depth and the zero-crossing overturning depth, relative to the AMOC depth. As such, negative values of deep overturning transports are associated with southward flows and a positive linear trend represent a decline in transport. The simulated AMOC transport is also sign‑reversed when estimating its trend, ensuring direct comparability with the deep overturning transport.

\section*{AMOC and opposing boundary trends in deep overturning}

\begin{figure}
    \centering
    \includegraphics[height=0.68\textheight]{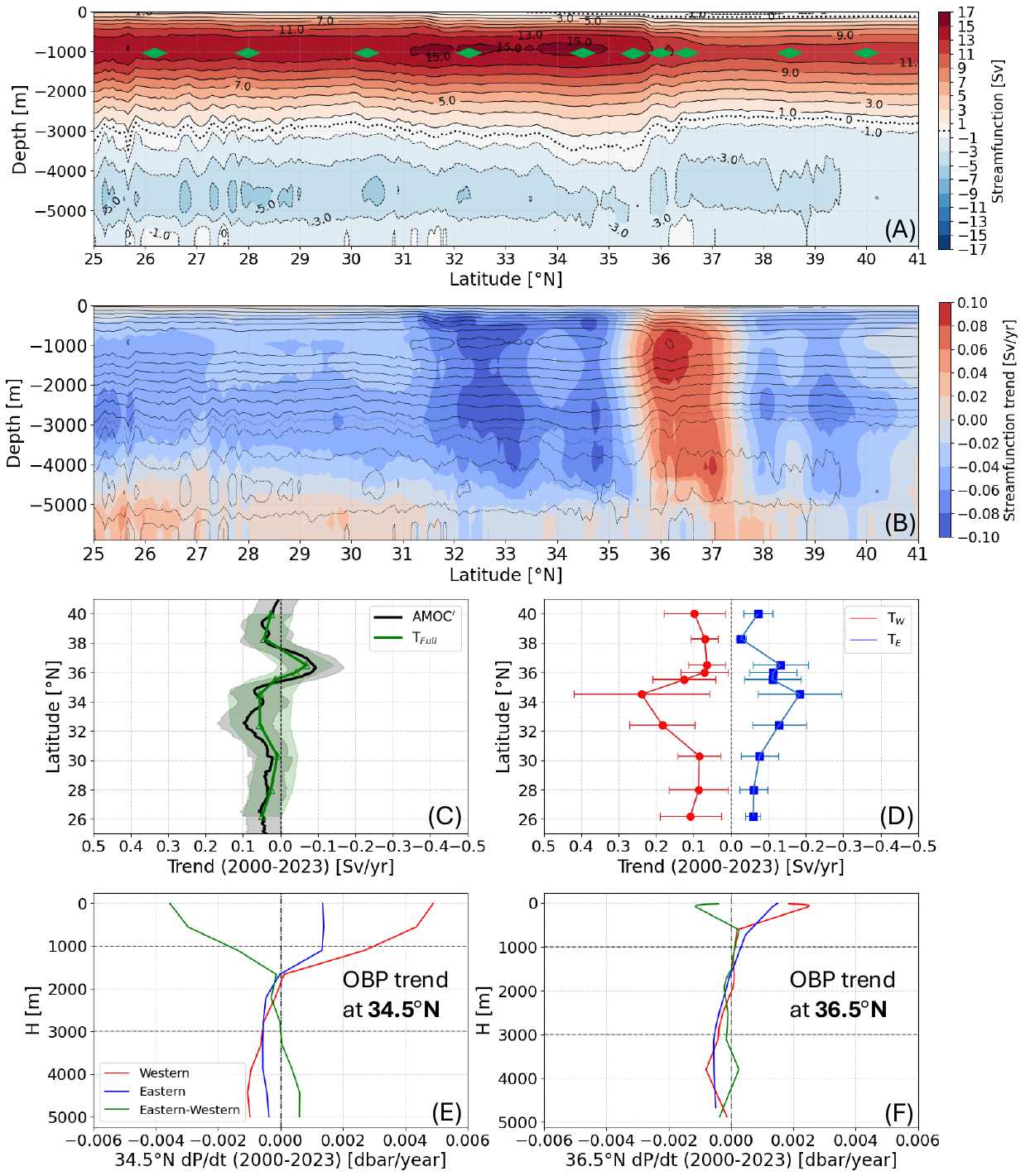}
    \caption{\textbf{AMOC and the opposing trends in deep western and eastern overturning transports in the mid-latitude North Atlantic.} (A) Temporal mean meridional overturning stream function computed by zonal integration of velocity fields across the mid-latitude North Atlantic basin. (B) Colors indicate the linear trend in the overturning stream function, and contours show the mean overturning stream function from panel (A). (C) Linear trends with uncertainties of 95\% significance level (shading) in deep full overturning transport (green; $T_{Full}$) at ten selected latitudes and depths (green rhombuses in the panel A) and AMOC anomalies (black) throughout the basin. (D) Linear trends with uncertainties (error bars) in deep western overturning transport (red; $T_{W}$) and deep eastern overturning transport (blue; $T_{E}$) at the ten selected latitudes. (E, F) Vertical profiles of OBP trends at the western boundary, eastern boundary, and their difference for the 34.5$^{\circ}$N and 36.5$^{\circ}$N section. }\label{Fig1}
\end{figure}

Within the mid-latitude North Atlantic, the zonally integrated and time mean meridional overturning streamfunction in our model exhibits several notable features (Fig.~\ref{Fig1}A). The AMOC depth remains relatively constant at around 1000 m, whereas the zero-crossing overturning depth, marking the boundary between the upper and lower overturning cells, is around 3400 m south of 35.5$^{\circ}$N and becomes shallower, around 2700 m, to the north of that latitude. The strength of the overturning maximum also exhibits weaker values north of 35.5$^{\circ}$N, with its strongest values occurring between 32$^{\circ}$N and 35$^{\circ}$N. The upper overturning cell shows a widespread negative trend across most latitudes, featuring pronounced values near 32–35$^{\circ}$N, except for a localized positive trend between 35.5$^{\circ}$N and 37.5$^{\circ}$N (Fig.~\ref{Fig1}B). Consistently, the basin-wide AMOC trend during 2000–2023 displays a similar meridional structure, reflecting the trend pattern of the upper overturning cell (Fig.~\ref{Fig1}C). In addition, positive trend values are present in the lower part of the lower overturning cells (below 4500 m, Fig.~\ref{Fig1}B), corresponding to a weakening of the abyssal overturning.\\

The primary climatic influence of the AMOC stems from its role in oceanic meridional heat transport (MHT), with variations in MHT directly influencing regional to global temperature patterns \cite{johns2023towards}. The tight coupling between MHT and AMOC is well established at 26.5$^{\circ}$N, where RAPID array observations demonstrate a strong correlation between the two \cite{johns2011continuous,johns2023towards}, implying that large-scale AMOC changes are accompanied by corresponding MHT adjustments at that latitude. In our model, this linear relationship between AMOC and MHT holds broadly (Figure~\ref{figS1}): their correlations exceed 0.7 across the basin and remain statistically significant at all latitudes of the domain considered. However, a noticeable weakening in correlation is observed between 35$^{\circ}$N and 38$^{\circ}$N, implying a transitional zone where the contribution of overturning to the MHT is probably different to adjacent regions. The AMOC and MHT trends show consistent sign alignment across latitudes. A positive trend (weakening) dominates most latitudes, except between 35.5$^{\circ}$N and 37.5$^{\circ}$N where MHT also exhibits negative trends (strengthening). North of 39$^{\circ}$N, the MHT weakening trend persists, while the AMOC trend decays to near zero by 41$^{\circ}$N. This indicates that in the subpolar region, changes in MHT are no longer dominated by the AMOC but may instead be governed by horizontal circulation, such as the Gulf Stream extension \cite{lozier2019sea}.\\

At 26.5$^{\circ}$N, the simulated AMOC compares reasonably with observations, though with some systematic differences (Figure~\ref{figS2}). The simulated AMOC has a mean strength of 13.9 $\pm$ 2.8 Sv (mean ± standard deviation) over 2004–2023. Both the mean and variability are weaker than observed AMOC, compared with 17.1 $\pm$ 4.6 Sv reported by the RAPID array \cite{moat2024atlantic}. The time-mean AMOC depth (946 m) and the zero-crossing overturning depth (3324 m) are also shallower than the corresponding RAPID estimates (993 m and 4300 m, respectively). Such a relatively shallow zero-crossing overturning depth is typical of CMIP-class climate models, with nearly all models exhibiting a shallower overturning transition depth compared with RAPID observations \cite{bonan2025observational}. Many CMIP6 models also simulate an abyssal overturning cell associated with Antarctic Bottom Water (AABW) formation stronger than observed, a feature that is likewise present in our model. Consistent with the behavior of most CMIP6 models \cite{bonan2025observational}, which exhibit a strong relationship between present-day AMOC strength and its projected weakening under warming scenarios, the weaker AMOC in our model is accompanied by a weaker declining trend (0.06 $\pm$ 0.06 Sv/yr, 95\% confidence interval) compared with the observed RAPID trend of 0.09 $\pm$ 0.08 Sv/yr (2-$\sigma$ confidence intervals) over the same period \cite{moat2024atlantic}. \\

To estimate deep overturning transports in our model, integrated from the AMOC depth down to zero-crossing overturning depth, we extract continental slope OBP fields within a depth range of 1000-3000 m (see Materials and Methods) for both western (Figure~\ref{figS3}) and eastern boundaries (Figure~\ref{figS4}) of the mid-latitude North Atlantic (Fig.~\ref{Fig2}A). In the model, the AMOC depth is approximately 1000 m, whereas we adopt a fixed zero‑crossing overturning depth of 3000 m across the entire basin. Because estimating deep overturning transport from OBP cross-slope gradients along the full continental boundaries is computationally demanding, we focus on ten representative latitudes across the domain considered to compute deep overturning transports for the western boundary contribution ($T_{W}$), eastern boundary contribution ($T_{E}$), and the combined contributions from $T_{W}$ and $T_{E}$ ($T_{Full}$), see Equations \ref{eq4}, \ref{eq5}, and \ref{eq6} in Materials and Methods. More latitudes are sampled near 36$^{\circ}$N, which appears to be a transition region. From all latitudes considered,  we identify a meridionally consistent pattern of a declining trend for $T_{W}$ and a strengthening trend for $T_{E}$ (Fig.~\ref{Fig1}D). The magnitude of these opposing trends peak at around 32-35$^{\circ}$N, aligned with the latitude of peak AMOC strength and the strongest declining trend in the AMOC (Fig.~\ref{Fig1}A). In most cases, the $T_{W}$ trends exceed those of $T_{E}$, resulting in a declining $T_{Full}$ trend across nearly all latitudes, except at 36$^{\circ}$N and 36.5$^{\circ}$N. The $T_{Full}$ trends at all representative latitudes closely track the AMOC trends (Fig.~\ref{Fig1}C). These results confirm that the trend in deep overturning transport derived from OBP cross-slope gradient provides a robust measure of the AMOC trend across the mid-latitude North Atlantic, with contrasting contributions from the western and eastern boundaries giving rise to the opposing trend patterns.

\section*{Opposite-sign vertical structure of the OBP trend}

The deep overturning transport is governed by the cross-slope OBP gradients at the western and eastern boundaries (see Equation~\ref{eq3} in Materials and Methods). Accordingly, the trends in $T_{W}$ and $T_{E}$ are expected to be closely linked to the vertical structure of OBP trends along the continental slopes of their respective boundaries. The trends in deep overturning transport exhibit a distinct meridional pattern (Fig.~\ref{Fig1}, C and D), with $T_{\mathrm{Full}}$ weakening at 32–35$^{\circ}$N and strengthening at 35.5–37.5$^{\circ}$N, accompanied by correspondingly larger and smaller magnitudes in the $T_{W}$ and $T_{E}$ trends in these two regions. We therefore select 34.5$^{\circ}$N and 36.5$^{\circ}$N from these two regions as representative latitudes to further examine how the vertical structure of the OBP trend influences the overturning transport trends. Both the western and eastern boundaries at 34.5$^{\circ}$N and 36.5$^{\circ}$N exhibit an opposite-sign vertical structure in the OBP trend: a positive trend above 1500–2000 m depth and a negative trend below (Fig.~\ref{Fig1}, E and F). A key distinction is that at 34.5$^{\circ}$N, the magnitude of the western OBP trend for the full depth generally exceeds that of the eastern boundary, whereas at 36.5$^{\circ}$N, the western trend becomes comparable to, or even weaker than, its eastern counterpart. This contrast reflects the relative magnitudes of the $T_{W}$ and $T_{E}$ trends at these latitudes. These results illustrate the link between the overturning transport trend and the vertical structure of the OBP trend at these two latitudes. Therefore, we examine next whether this vertical structure is a robust feature across the broader mid-latitude North Atlantic.\\
\begin{figure}[htbp]  
    \centering
    \includegraphics[width=1\textwidth]{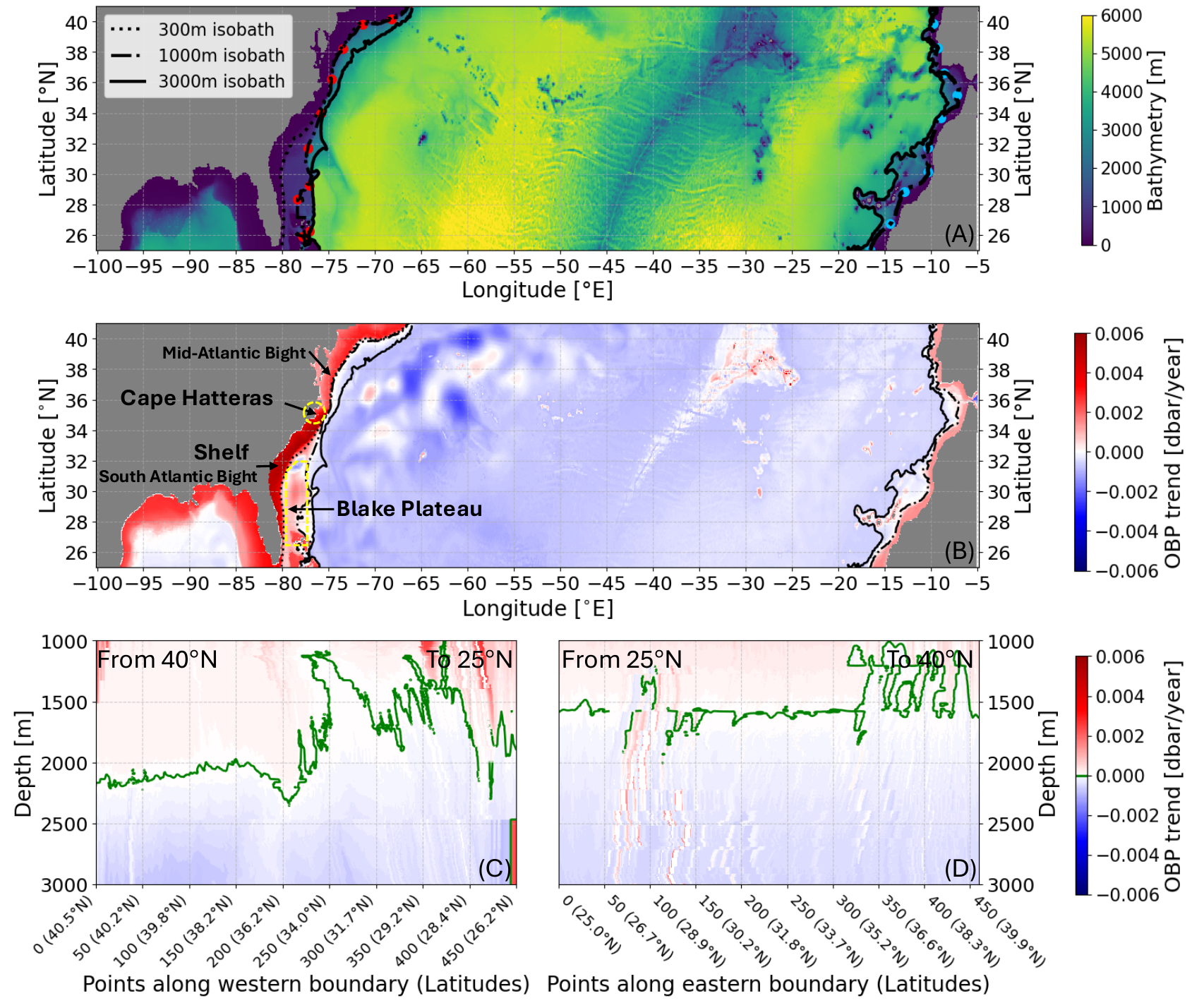}
    \caption{\textbf{The opposite-sign structure of the OBP trend across the mid-latitude North Atlantic.} (A) Bathymetry of the mid-latitude North Atlantic between 25$^{\circ}$N and 41$^{\circ}$N. Red and blue dots mark every 50 points along the traced isobaths on the western and eastern boundaries, respectively. Note that tracing begins at 41$^{\circ}$N on the western boundary and at 25$^{\circ}$N on the eastern boundary. (B) Horizontal distribution of the OBP trend across the mid-latitude North Atlantic. Yellow circle and box at the western boundary mark the approximate locations of Cape Hatteras and the Blake Plateau. (C, D) Vertical structure of the OBP trend at the western and eastern boundaries, shown in along-boundary versus depth coordinates. The smoothed green contour highlights the zero-OBP trend line. For clarity, this and the following figures include only the isobaths at the western and eastern continental boundaries and isobaths in other regions (e.g., the Gulf of Mexico and the Mid-Atlantic Ridge) are excluded. }\label{Fig2}
\end{figure}

Fig.~\ref{Fig2}B shows the horizontal distribution of the OBP trend throughout the mid-latitude North Atlantic basin. In general, the OBP exhibits positive trend values on both western and eastern continental shelves, and negative trend values in the ocean interior. To examine further the structure of the OBP trend with depth, we trace bathymetry contours in the 1000-3000 m depth range, and we extract continental slope OBP fields for both western and eastern boundaries so that we can plot OBP trend in an along-boundary versus depth coordinate system (Fig.~\ref{Fig2}, C and D). Such representation of the OBP trend reveals a meridionally consistent vertical (cross-slope) pattern, with positive values at shallower depths and negative values at deeper depths at both the western and eastern boundaries. At the western boundary, the transition between positive and negative OBP trends is substantially deep, around 2300 m, at latitude of around 36.2$^{\circ}$N (point 200). At latitudes near 34$^{\circ}$N (point 250), the transition depth sharply shoals to about 1200 m, and then deepens again to vary between 1500 and 2000 m to south of 31.7$^{\circ}$N (Fig.~\ref{Fig2}C). At the eastern boundary, aside from a localized region of sharp changes (points 50–150) associated with complex topography, the depth of the zero-crossing OBP trend remains relatively stable at approximately 1600 m except for a shoaling to north of latitude near 36$^{\circ}$N. Overall, the OBP trend exhibits a coherent cross-slope vertical structure at both boundaries, with a depth-dependent sign reversal that varies meridionally, showing stronger spatial variability at the western boundary and a comparatively stable structure at the eastern boundary. \\

Within the 1000–3000 m layer, we therefore observe a meridionally consistent opposite-sign (positive-up, negative-down) vertical structure of the OBP trend at both boundaries. This structure is associated with the widespread declining trend in $T_{W}$ and the strengthening trend in $T_{E}$. In this deep layer, the deep negative OBP trend contributes to a negative trend in the OBP cross-slope gradient. According to geostrophy (Equation~\ref{eq2} in Materials and Methods), a negative trend in the cross-slope OBP gradient produces a negative trend in the shear of the eastern boundary overturning contribution and a positive trend in the shear of the western boundary contribution. After performing the vertical integrations, this shear tendency results in a positive (declining) trend in $T_{W}$ at the western boundary and a negative (strengthening) trend in $T_{E}$ at the eastern boundary. By the same argument, a positive OBP trend in the upper part of this deep layer in turn leads to a strengthening trend in $T_{W}$ and a weakening trend in $T_{E}$.\\

The positive-up and negative-down OBP trends within the 1000–3000 m layer induce opposite contributions to the overall trend in deep overturning transports. This implies that the relative depth of the zero-crossing of the OBP trend within the 1000–3000 m layer at each boundary controls the magnitudes of the declining trend in $T_{W}$ and strengthening trend in $T_{E}$, respectively. A deeper zero-crossing depth corresponds to a weaker declining trend in $T_{W}$ and a weaker strengthening trend in $T_{E}$. Therefore, at the western boundary, the weakest declining trend in $T_{W}$ occurs near 36$^{\circ}$N, where the zero-crossing depth is the deepest, while the strongest declining trend occurs near 32-35$^{\circ}$N with the shallowest zero-crossing depth (Fig.~\ref{Fig2}C). In the same way, at the eastern boundary, the shallower zero-crossing depth north of 36$^{\circ}$N corresponds to the stronger strengthening trend in $T_{E}$ (Fig.~\ref{Fig2}D). Its magnitude becomes larger than that of $T_{W}$, resulting in a net strengthening trend in $T_{Full}$ at 35.5–37.5$^{\circ}$N.\\

In the simulated mid-latitude North Atlantic, we identify meridionally coherent but opposing overturning trends at the western and eastern boundaries, along with the strongest decline in the AMOC between 32$^{\circ}$ and 35$^{\circ}$N and a localized strengthening between 35.5$^{\circ}$ and 37.5$^{\circ}$N. These features share a common origin: they arise from the vertical structure of the trend in the OBP cross-slope gradient, and therefore from the spatial structure of the OBP trend itself. We now turn to identifying the physical mechanisms, under climate change as represented in the model, that give rise to this spatial structure of the OBP trend in the mid-latitude North Atlantic.

\section*{What drives the OBP trend structure under climate change?}

According to the calculation of the OBP in our model (see Equation~\ref{e1} in the Materials and Methods), the temporal tendency of the OBP can be decomposed into two contributions as follows:
\begin{equation}
\frac{\partial P'(z = -H, t)}{\partial t} = g\rho_{0}\frac{\partial \left[ \int_{z = -H}^{z = \eta(t)} \frac{\rho(z, t)-\rho_{0}}{\rho_{0}}\, dz \right]}{\partial t} + g\rho_{0}\frac{\partial \eta(t) }{\partial t},
\label{e4}
\end{equation}
where the first term on the right-hand side represents the contribution from full-depth density changes (hereafter density term) and the second term represents the contribution from mean sea level changes (herefafter sea level term). We examine how variations in these two terms act together to produce the characteristic spatial structure observed in the 2000-2023 OBP linear trend in the mid-latitude North Atlantic.\\
\begin{figure}[htbp]  
    \centering
    \includegraphics[width=1\textwidth]{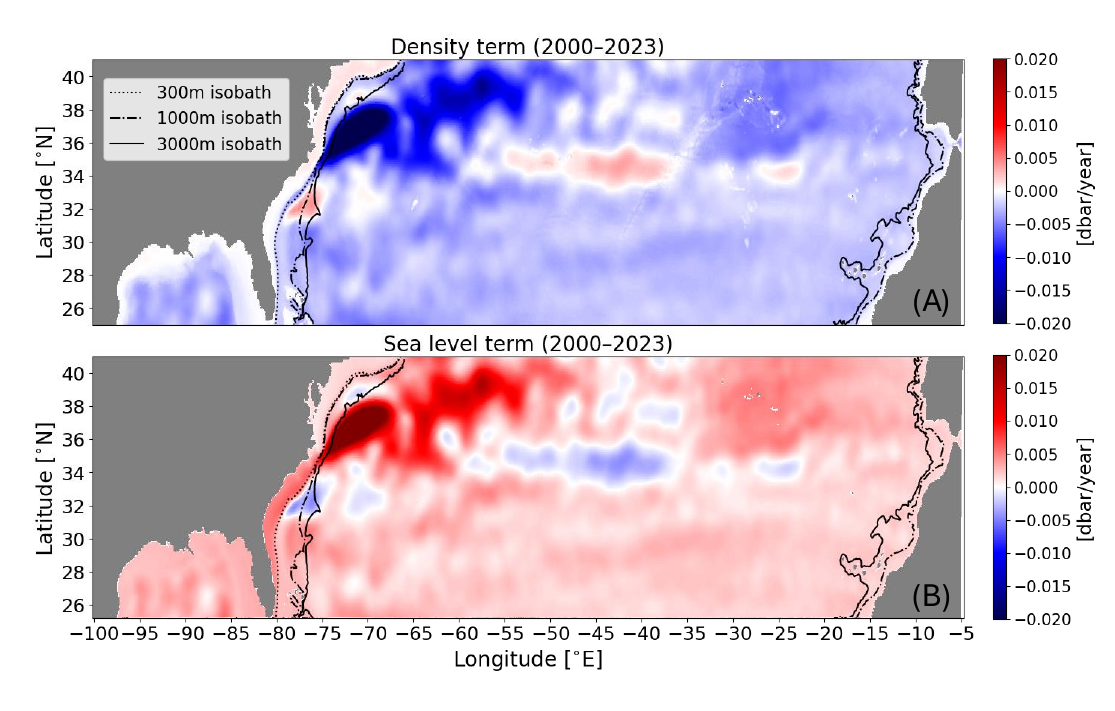}
    \caption{\textbf{The OBP trend is decomposed into contributions from the trend in ocean water density and sea level changes} (A and B) Horizontal distribution of the density term and sea level term in the mid-latitude North Atlantic during 2000-2023. Note that the two panels in this figure use a consistent color scheme with Fig.~\ref{Fig2}B but different scales, as the density and sea level terms are an order of magnitude larger than the OBP trend.}\label{Fig3}
\end{figure}

As shown in Fig.~\ref{Fig3}A, the density term is generally negative across the ocean interior, with pronounced negative values over the Gulf Stream region (north of about 35$^{\circ}$N) and localized positive values to the south of the Gulf Stream (32-34$^{\circ}$N) and the middle of the basin (35$^{\circ}$N, 55$^{\circ}$W-35$^{\circ}$W). In contrast, at the western boundary in waters shallower than 300 m, the density term contributes a weak positive signal to the north of Cape Hatteras (Mid-Atlantic Bight, see Fig.~\ref{Fig2}B) and near-zero signal to the south (Fig.~\ref{Fig3}A). The sea level term has a similar distribution pattern as the density term overall, but with an opposing sign in the ocean interior, showing how most of this common pattern reflects steric changes to the sea level (Fig.~\ref{Fig3}B). However, at the western boundary, the sea level term is of the same positive sign as the density term north of Cape Hatteras within the Mid-Atlantic Bight, resulting in a net positive OBP trend there. To the south of Cape Hatteras and inshore of the 300 m isobath, the density term is near zero but the sea level term is positive and relatively stronger than to the north, resulting in a net positive strong OBP trend in the South Atlantic Bight. At the eastern boundary, the sea level term is uniformly positive from offshore to inshore irrespective of the shallowing isobaths, whereas the density term transitions from negative in the interior and offshore to near zero within a narrow coastal strip, resulting in a positive OBP trend inshore of approximately the 1000 m isobath. Therefore, in the model, the effects of decreasing density and rising sea level on the OBP trend do not exactly cancel each other through steric adjustment. Instead, their imbalance causes the spatial pattern of OBP trend across much of the basin (Fig.~\ref{Fig2}B). Specifically, sea level rise dominates the positive OBP trend at shallower depths on the continental shelves, whereas density decrease predominantly drives the negative OBP trend at deeper depth on the continental slope and in the ocean interior. 
Thus, in the warming climate simulated by the model over 2000–2023 (as shown in Figures~\ref{figS5} and \ref{figS6}), the combined effects of sea level rise and decreasing ocean density shape the spatial structure of the OBP trend. In particular, they explain the meridional variations in the depth of the OBP trend zero-crossing at the basin boundaries (Fig.~\ref{Fig2}, C and D). These boundary OBP trends form a meridionally coherent pattern that reflects the boundary opposing contributions to Atlantic overturning and leads to a weakening AMOC throughout the mid-latitude North Atlantic.\\

\begin{figure}[htbp]  
    \centering
    \includegraphics[width=1\textwidth]{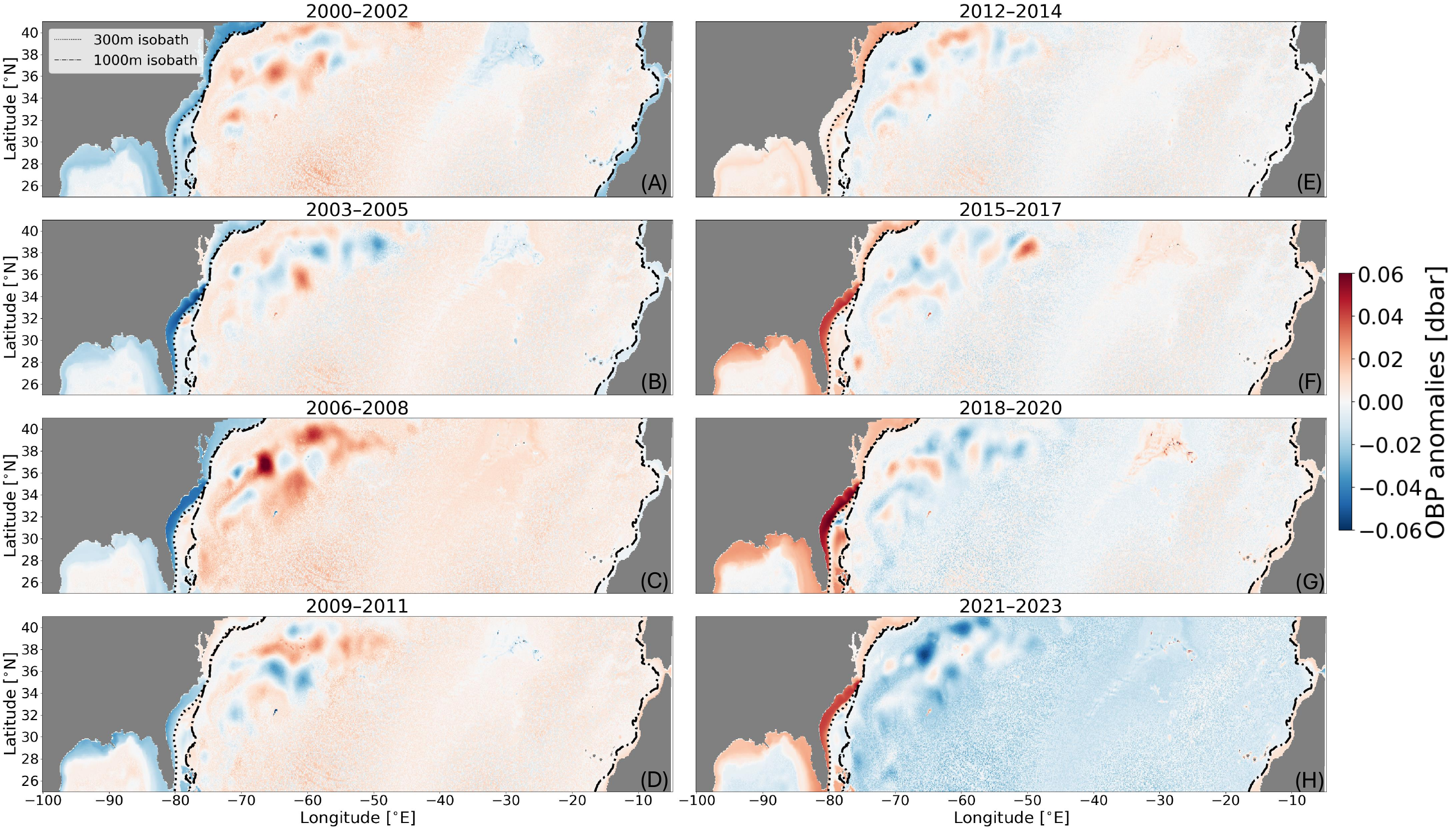}
    \caption{\textbf{Evolution of OBP anomalies in the mid-latitude North Atlantic during 2000–2023.} (A-H) OBP anomalies with respect to the 2000-2023 time-mean in consecutive 3-year intervals, chosen to reduce interannual variability while retaining sufficient temporal resolution to illustrate the gradual evolution of the anomalies.}\label{Fig4}
\end{figure}

To understand how the 2000-2023 OBP trend pattern arise, we examine the spatial evolution of OBP anomalies over 2000–2023 (Fig.~\ref{Fig4}), which generally shows an increase along the western and eastern boundaries and a decrease in the ocean interior, consistent with the spatial pattern of the OBP trend shown in Fig.~\ref{Fig2}B. However, the temporal evolution of these anomalies further reveals that OBP variability along the western and eastern boundaries exhibits distinct behaviors, which are not apparent from the spatial distribution of the long-term trend alone. OBP anomalies at the eastern boundary are primarily driven by sea level changes that are spatially uniform both inshore and offshore, showing little dependence on bathymetric contours and latitude (Figure~\ref{figS7}). In contrast, OBP anomalies at the western boundary are closely constrained by isobaths, with Cape Hatteras serving as a key separation point. From 41$^{\circ}$N to near Cape Hatteras, most OBP anomalies at the western boundary before 2012 are also driven by sea level changes (Figure~\ref{figS7}). In contrast, density changes typically lead to weak positive anomalies during 2012-2014, 2015-2017, and 2021-2023 (Figure~\ref{figS8}, B, D, and F), associated with enhanced increased salinity in the salty North Atlantic Mode Waters \cite{hogikyan2024hydrological}. South of Cape Hatteras, stronger OBP anomalies develop on the South Atlantic Bight, where sea level changes always dominate during 2000-2023 (Figure~\ref{figS7}). During 2015–2023, pronounced positive OBP anomalies associated with sea level changes propagate from the South Atlantic Bight toward Cape Hatteras (Fig.~\ref{figS7}F–H). These sea level changes arise from variations in Gulf Stream transport, owing to the large cross-stream sea level gradients of the Gulf Stream and its proximity to the shelf \cite{ezer2019regional}. The positive OBP anomalies near Cape Hatteras are dynamically linked to the open ocean through Gulf Stream variability \cite{wu2025gulf}. \\

Overall, at the western boundary, OBP anomalies generally weaken from 41$^{\circ}$N toward Cape Hatteras regardless of whether they are driven by density changes or sea level changes, whereas intensified continental shelf-origin anomalies driven by sea level changes on the South Atlantic Bight extend to near Cape Hatteras. This results in the weakest declining trend and the strongest declining trend in the overturning transports derived from OBP cross-slope gradient to the north and south of Cape Hatteras, respectively. In addition, OBP anomalies transitioning from negative to positive values are generally stronger at the western boundary relative to the eastern boundary at the same latitude (Fig.~\ref{Fig4}). Consequently, the magnitude of the declining trend in western boundary overturning is generally larger than the magnitude of the strengthening trend at the eastern boundary, producing a widespread declining trend in the AMOC across most latitudes of the mid-latitude North Atlantic. 

\section*{Discussion}

This study employs a high-resolution model to show that the interplay between rising sea level and decreasing density under a warming climate generates an opposite-sign spatial structure of the ocean bottom pressure (OBP) trend at both the western and eastern boundaries of the mid-latitude North Atlantic. This structure, in turn, produces opposing trends in deep overturning transport linked to OBP cross-slope gradients at the western and eastern boundaries through the zonally integrated meridional geostrophic balance. Since the declining trend in western boundary overturning is generally stronger than the strengthening trend in the east, the model yields a meridionally consistent decline in the deep overturning transport across the full section, which closely captures the declining trend in the AMOC throughout the mid-latitude North Atlantic. These results are consistent with and provide a clear mechanistic explanation for the findings of a previous observational study \cite{xing2026meridionally}. Together, they advocate for the inclusion of OBP measurements in future AMOC observing systems, despite the current limitations in AMOC observational coverage and the technical constraints of bottom pressure recorders \cite{mccarthy2020sustainable,hughes2018window,harmon2026implications}.\\

The depths that primarily define the vertical structure of the overturning circulation in the North Atlantic (AMOC depth and zero-cross overturning depth) show changes that are linked to the long-term AMOC trend, yet this linkage  has received limited attention. The AMOC trends in our model are weaker than those inferred at the RAPID array. Several factors may contribute to this difference, including the weaker simulated AMOC strength and the shallower zero-crossing overturning depth. A recent study has shown that climate models with stronger and deeper present-day overturning tend to project larger weakening and shoaling under global warming, because a less stratified North Atlantic allows surface buoyancy flux anomalies to penetrate deeper, producing larger density changes at depth and consequently a larger AMOC decline \cite{bonan2025observational}. Recent RAPID observations indicate that both the AMOC depth and the zero-crossing overturning depth have shoaled over the past two decades \cite{moat2024atlantic,xing2026meridionally}. Because the deep overturning transport is calculated within these two depth boundaries, such changes in the vertical structure of the overturning can substantially influence the magnitude of its long-term trend, and thus the estimated AMOC trend. This highlights the importance of carefully selecting and regularly reassessing the depth ranges used to estimate deep overturning transport. Moreover, we argue that future AMOC observing system designs should more explicitly monitor how these key depths evolve in response to overturning changes.\\

This model study identifies a special transition region between 32 and 37.5$^{\circ}$N, with 35.5$^{\circ}$N as a key boundary for the AMOC strength (Fig.~\ref{Fig1}A) as well as for the trends in AMOC and deep overturning transport (Fig.~\ref{Fig1}, B and C). At the western boundary, Cape Hatteras, located near 35.5$^{\circ}$N, marks the separation of the Gulf Stream from the continental margin into the ocean interior. This separation favors the shallowest zero-crossing depth of the OBP trend due to the strongest negative sea level term between 1000 and 3000 m (Fig.~\ref{Fig2}C) at 32-35$^{\circ}$N. These features give rise to the strongest declining trends in western boundary overturning transport (Fig.~\ref{Fig1}D), and therefore the largest AMOC declining trend at 32-35$^{\circ}$N. Further north (35.5-37.5$^{\circ}$N), the deepening zero-crossing depth of the OBP trend due to the strongest positive sea level term below 1000 m corresponds to much weaker declines in western boundary overturning. In contrast, the shoaling zero-crossing depth of the OBP trend at the eastern boundary induce stronger strengthening in eastern boundary overturning. This leads to a net strengthening of deep overturning transport, and thus a strengthening of the AMOC at 35.5-37.5$^{\circ}$N. In addition, in the framework of the transformed Eulerian mean, the residual meridional circulation is defined as the sum of the Ekman transport and the eddy-induced transport, which together set the overturning structure and stratification \cite{karsten2002role,marshall2003residual,badin2010buoyancy}. Although this theory was originally developed for the Antarctic Circumpolar Current region, it is expected to be applicable to other baroclinically active western boundary current systems, including the Gulf Stream. From an energetics perspective, the positive eddy kinetic energy (EKE) trend near Cape Hatteras (Figure~\ref{figS9}) reflects a strengthening of localized eddy-induced transport. Given that the Ekman transport exhibits no significant trend over the study period, the enhanced eddy-induced transport leads to a strengthened residual circulation and thus a positive AMOC trend at 35.5-37.5$^{\circ}$N. The sign-contrast in AMOC trends observed in this region leads to a northward displacement of the AMOC streamfunction, most clearly seen in the significant positive linear trend of the maximum AMOC latitude (Figure~\ref{figS10}A). This finding suggests that the AMOC response to climate change is more nuanced than a simple, meridionally coherent slowdown or acceleration. Rather, it is characterized by a spatial reorganization of its circulation, involving distinct regional changes in dynamics.  \\


Rather than exhibiting a basin-wide decline, the modeled overturning circulation displays spatial heterogeneity in its long-term evolution across the North Atlantic. The contrasting strengthening observed at 35.5–37.5$^{\circ}$N is one example. In addition, a positive trend in overturning streamfunction appears south of 32$^{\circ}$N below about 4800 m (Fig.~\ref{Fig1}B). Our model shows decreasing temperature and Ocean Heat Content (OHC) trends (Figure~\ref{figS5}C and Figure~\ref{figS6}D), together with the increasing density trend at these depths (Figure~\ref{figS5}D), indicate that the abyssal water in the mid-latitude North Atlantic is becoming colder and denser due to the cooling of North Atlantic Deep Water coming from the north \cite{johnson2024refined}. This abyssal cooling increases deep-ocean stratification, contributing to the positive trend in overturning transport in the abyssal ocean. As our model reasonably reproduces observed abyssal climate changes (e.g., temperature and OHC trends) consistent with long-term observations \cite{johnson2024refined,desbruyeres2016deep}, we are confident that it captures the basin-wide and abyssal adjustment of the overturning circulation to climate change. Our simulated results imply that AMOC observations at a single latitude, while capable of detecting large-scale signals and local variability, may be insufficient and potentially misleading for characterizing the full basin-scale adjustment of the overturning circulation.\\

The high-resolution ocean model used in this study helps overcome several observational limitations, including the short and non-uniform duration of available records and the sparse distribution of observational arrays that cannot fully cover the basin. The model enables a robust diagnosis of the relationship between OBP-based metrics (including OBP trends and OBP cross-slope gradients trends) and overturning transport trends. This relationship reflects density and sea level changes associated with climate warming that generate opposite-sign spatial structures in OBP trends, thereby inducing opposing boundary contributions to overturning transport, as observed by the RAPID array \cite{xing2026meridionally}, ultimately leading to the net decline in the AMOC. Although the simulated AMOC is weaker and shallower than observed, resulting in a reduced AMOC trend magnitude, we are confident that the mechanistic insights derived from the OBP framework remain applicable for interpreting overturning transport trends inferred from observations. At the same time, our results highlight that trend estimates derived from a single observational array should be interpreted with caution, and emphasize the importance of monitoring the overturning circulation at multiple latitudes, with observational analyses evaluated in conjunction with model simulations.

\section*{Methods}

\subsubsection*{Eddy-rich double-nested model}
The numerical simulations analyzed here are conducted using NEMO version 4.2.2 \cite{madec2023nemo}, coupled to the SI3 sea-ice model \cite{vancoppenolle2023si3}. The model setup follows the general design of the global 1/4$^{\circ}$ GOSI9 configuration developed jointly by the National Oceanography Centre and the UK Met Office within the Joint Marine Modelling Programme \cite{guiavarc2025gosi9}. Enhanced horizontal resolution is achieved through a two-level AGRIF nesting strategy in the subtropical North Atlantic. The first nest refines the grid to 1/12$^{\circ}$ over 6–42$^{\circ}$N, while a second nest further increases the resolution to 1/36$^{\circ}$ between 19–31$^{\circ}$N. The vertical discretization consists of 75 levels, with layer thicknesses increasing from 1 m near the surface to approximately 250 m at depth, and bottom topography represented using partial-steps. Atmospheric forcing is prescribed using surface fluxes derived from the Japanese Meteorological Agency reanalysis JRA55-do \cite{tsujino2018jra}. The simulation spans the period 1976–2023. For the present analysis, we focus on the mid-latitude North Atlantic (25–41$^{\circ}$N), a region well suited for diagnosing basin-scale AMOC variability and continental-slope ocean bottom pressure. Monthly model output at 1/12$^{\circ}$ resolution is used throughout this domain. To facilitate comparisons with observations from long-term mooring arrays, including RAPID \cite{moat2024atlantic} and Line W \cite{lebras2023atlantic}, and to minimize the influence of model spin-up, we restrict our analysis to the 2000–2023 period.\\

We present diagnostics which show the background evolving climatic conditions in the North Atlantic study region over model years 2000-2023. As shown in Figure~\ref{figS5}, both sea surface temperature (SST) and sea surface height (SSH) exhibit basin-wide positive trends throughout the mid-latitude North Atlantic. Figure~\ref{figS6} further displays the horizontal patterns of depth-integrated ocean heat content (OHC) trends across four layers (0–700 m, 700–2000 m, 2000–4000 m, and $>$4000 m). In the upper 4000 m, OHC trends are predominantly positive, with stronger positive trend in the western half basin and weaker or localized negative trends in the eastern basin. In contrast, the abyssal layer below 4000 m is characterized by widespread negative OHC trends. Overall, the features demonstrate that the model simulation during 2000–2023 captures a warming climate state that is broadly consistent with observation-based datasets and studies \cite{karnauskas2021atmospheric,prandi2021local,liao2022comparative,li2023recent,johnson2024refined,desbruyeres2016deep}.

\subsection*{OBP calculation in the model}

Our model computes ocean bottom pressure (OBP) based on the hydrostatic balance following the approach adopted by many others (e.g., CMIP6 \cite{liu2025assessment}). Considering a water column of the mid-latitude North Atlantic (Figure~\ref{figS11}), the OBP anomaly $P'(z = -H, t)$ (also denoted as OBP) at depth $z=-H$ and time $t$ can be expressed as:
\begin{equation}
P'(z = -H, t) = g\rho_{0} \left[ \int_{z = -H}^{z = \eta(t)} \frac{\rho(z, t)-\rho_{0}}{\rho_{0}} dz + \eta(t) \right],
\label{e1}
\end{equation}
where $\rho(z,t)$ is the in-situ density of seawater, $g$ is the gravitational acceleration, $\eta$ denotes the time-varying sea surface height relative to the geoid ($z=0$), and $\rho_0$ is the reference density. Although atmospheric pressure contributes to OBP in the real ocean, it is not supplied as part of the atmospheric forcing for our ocean model configuration and is therefore not present in this calculation.

\subsection*{Measuring deep overturning transport}

Measuring deep overturning transport using the stepping method \cite{hughes2013test} has been described in detail in our previous work \cite{xing2026meridionally}. Here, we provide a brief summary of the method used to estimate deep overturning transport from cross-slope gradients of OBP. Starting from the zonal component of the horizontal geostrophic momentum balance, we obtain the expression
\begin{equation}
\label{eq1}
Q(y,z) \equiv \int_{x_W}^{x_E} v_g dx = \frac{1}{\rho f}\left[p_E(y,z) - p_W(y,z)\right],
\end{equation}
which states that $Q(y,z)$, the meridional geostrophic transport per unit depth at latitude $y$, is proportional to the difference in OBP between the eastern and western continental slopes.\\

Our interest lies in overturning transports, associated with vertical shear and compensating flows at different depths \cite{bingham2008determining}. The vertically integrated boundary OBP gradient has been shown to be indicative of overturning processes in the North Atlantic MOC. This expectation was confirmed by further study \cite{elipot2014observed} where transport profiles together with independent bottom pressure recorder (BPR) measurements from the first year of the RAPID array are used to demonstrate that the boundary pressure gradient is proportional to the vertical shear of the geostrophic transport profile. Because the spatial distribution of individual mooring arrays makes it impossible to obtain continuous OBP measurements along the continental slope, direct observations of the along-slope OBP structure are not available. However, following Elipot et al. \cite{elipot2017observed} and Xing et al. \cite{xing2026meridionally}, a stepping method can be employed, in which the cross-slope OBP gradient is inferred from density and along-slope velocity measurements, allowing the overturning transport to be estimated. Consistent with these observational studies, we therefore continue to use the vertical structure of the cross-slope OBP gradient as the key diagnostic quantity for overturning variability. Therefore, the relevant quantity for diagnosing overturning is the vertical structure of the cross-slope OBP gradient. Taking the vertical derivative of (\ref{eq1}) yields
\begin{equation}
\label{eq2}
\frac{\partial Q(y,z)}{\partial z} = \frac{1}{\rho f}\frac{\partial}{\partial z}\left[p_E(y,z) - p_W(y,z)\right].
\end{equation}

Equation (\ref{eq2}) shows that the vertical shear of meridional transport, and thus the overturning circulation, can be decomposed into two contributions: one from the vertical gradient of eastern-boundary OBP and one from that of the western boundary. Each contribution can be vertically integrated to yield an overturning transport profile at the individual boundary \cite{hughes2013test,xing2026meridionally}.\\

In this study, we compute the deep full overturning transport $T_{Full}$ by vertically integrating the right-hand side of (\ref{eq2}) twice over the depth range bounded by the fixed depth of the overturning streamfunction maximum ($z_{AMOC}$) and the fixed depth of the zero crossing of the overturning streamfunction ($z_{b}$):
\begin{equation}
\label{eq3}
T_{\mathrm{Full}}(y) = \frac{1}{\rho f}
\int_{z_b}^{z_{\mathrm{AMOC}}}
\left[
\int_{z_b}^{z_{\mathrm{AMOC}}}
\frac{\partial}{\partial z'}
\left[p_E(y,z') - p_W(y,z')\right] dz'
\right] dz.
\end{equation}

We further separate this deep full overturning transport into eastern and western boundary contributions by integrating the OBP gradients at each boundary independently:
\begin{equation}
\label{eq4}
T_E(y) = \frac{1}{\rho f}
\int_{z_b}^{z_{\mathrm{AMOC}}}
\left[
\int_{z_b}^{z_{\mathrm{AMOC}}}
\frac{\partial p_E(y,z')}{\partial z'}dz'
\right] dz,
\end{equation}
\begin{equation}
\label{eq5}
T_W(y) = -\frac{1}{\rho f}
\int_{z_b}^{z_{\mathrm{AMOC}}}
\left[
\int_{z_b}^{z}
\frac{\partial p_W(y,z')}{\partial z'}dz'
\right] dz.
\end{equation}

The anomalies of deep full overturning transport is then given by the sum of the anomalies of two boundary contributions:
\begin{equation}
\label{eq6}
T_{Full}^{'}(y) = T_E^{'}(y) +T_W^{'}(y).
\end{equation}

\subsection*{Extracting continental slope bottom pressure}

Extracting continental-slope bottom pressure is essential for estimating the deep overturning transport derived from cross-slope gradients of OBP and for examining the vertical structure of OBP trends. The overall methodology was originally proposed by Hughes et al. \cite{hughes2018window}, where a comprehensive description can be found. For the purpose of extracting OBP and estimating deep overturning transport from our model output, we follow this approach with several adaptations tailored to our model configuration. The corresponding code is available on Zenodo (see Zenodo link in Data and materials availability), and a detailed description of the procedure, consisting of three main steps, is provided below.

\paragraph{1. Isobath tracing and construction of an isobath stack} We trace targeted isobaths from the model bathymetry using a contour-following algorithm and subsequently organize all traced contours into an ordered stack to ensure spatial continuity along the continental slope. The algorithm begins from a user-defined grid cell close to the desired contour value and identifies intersections between cell edges and the prescribed contour using sign-change detection. The contour is then traced in both directions using a marching-squares–type procedure, in which successive points are located through linear interpolation along intersected edges. Boundary conditions (periodic and non-periodic) are treated consistently, and tracing terminates when the contour closes or reaches an open boundary. This process yields an ordered sequence of grid coordinates representing a continuous isobath.

\paragraph{2. Identification of cross-slope points} For each isobath, we identify grid locations on the continental slope that are approximately aligned with the local cross-isobath (normal) direction. Because the isobath produced from step 1 can be noisy due to grid discretization, staircase artifacts, and projection effects, we first apply a hybrid smoothing procedure that combines adaptive moving-average filtering \cite{shan2022novel}, Gaussian convolution with reflective boundaries \cite{buades2005review}, and spline-based curve reconstruction subject to arc-length preservation. This approach effectively removes grid-scale noise while retaining large-scale curvature and contour geometry. Local tangent directions are then estimated via linear regression within a sliding window, and the corresponding normal vectors are defined as the perpendicular directions. To ensure stable normal estimates in regions with high curvature, we further apply an elliptical weighting scheme that aggregates information from a large neighborhood, favoring nearby points in the projected normal direction and suppressing along-isobath variations. For each reference point on a shallower isobath, deeper-isobath correspondences are determined by projecting along the estimated normal direction. Candidate matches are identified within a prescribed search radius and selected based on the smallest positive projection distance onto the normal vector. This procedure establishes consistent cross-isobath mappings that capture the geometric deformation of the continental slope from the upper slope to the deep basin.

\paragraph{3. Extraction of bottom pressure} Finally, OBP values are sampled at all selected cross-slope points for every time step. These extracted data form a three-dimensional array, spanning time, reference points along the boundary, and depth. It provides the basis for computing cross-slope OBP gradients as well as for examining the depth-dependent structure of OBP variability and long-term trends.

\section*{Data availability}
The model data outputs used in this study are available at \href{https://doi.org/10.5281/zenodo.14733917}{https://doi.org/10.5281/zenodo.14733917}. 
The python scripts for extracting ocean bottom pressure are available at \href{https://doi.org/10.5281/zenodo.17970821}{https://doi.org/10.5281/zeno\\do.17970821}.



\clearpage 

%
\bibliography{references} 

@Article{bryden2009adjustment,
AUTHOR = {Bryden, H. L. and Mujahid, A. and Cunningham, S. A. and Kanzow, T.},
TITLE = {{Adjustment of the basin-scale circulation at 26$^\circ$N to variations in Gulf Stream, deep western boundary current and Ekman transports as observed by the Rapid array}},
JOURNAL = {Ocean Science},
VOLUME = {5},
YEAR = {2009},
NUMBER = {4},
PAGES = {421--433},
URL = {https://os.copernicus.org/articles/5/421/2009/},
DOI = {10.5194/os-5-421-2009}
}

@article{marshall2003residual,
  title={Residual-mean solutions for the Antarctic Circumpolar Current and its associated overturning circulation},
  author={Marshall, John and Radko, Timour},
  journal={Journal of Physical Oceanography},
  volume={33},
  number={11},
  pages={2341--2354},
  year={2003}
}

@article{karsten2002role,
  title={The role of eddy transfer in setting the stratification and transport of a circumpolar current},
  author={Karsten, Richard and Jones, Helen and Marshall, John},
  journal={Journal of Physical Oceanography},
  volume={32},
  number={1},
  pages={39--54},
  year={2002}
}

@article{badin2010buoyancy,
  title={On the buoyancy forcing and residual circulation in the Southern Ocean: The feedback from Ekman and eddy transfer},
  author={Badin, Gualtiero and Williams, Richard G},
  journal={Journal of physical oceanography},
  volume={40},
  number={2},
  pages={295--310},
  year={2010}
}

@article{lozier2019sea,
  title={A sea change in our view of overturning in the subpolar North Atlantic},
  author={Lozier, M Susan and Li, Feili and Bacon, Sheldon and Bahr, F and Bower, Amy S and Cunningham, SA and de Jong, M Femke and de Steur, Laura and deYoung, Brad and Fischer, J{\"u}rgen and others},
  journal={Science},
  volume={363},
  number={6426},
  pages={516--521},
  year={2019},
  publisher={American Association for the Advancement of Science}
}

@article{liu2025assessment,
  title={Assessment of ocean bottom pressure variations in CMIP6 HighResMIP simulations},
  author={Liu, Le and Schindelegger, Michael and B{\"o}rger, Lara and Foth, Judith and Gou, Junyang},
  journal={Ocean Science},
  volume={21},
  number={5},
  pages={2149--2167},
  year={2025},
  publisher={Copernicus Publications G{\"o}ttingen, Germany}
}

@article{harmon2026implications,
  title={Implications for oceanographic and seafloor geodetic applications due to settling of self-calibrating bottom pressure recorders},
  author={Harmon, Nicholas and Rychert, Catherine A and Moat, Ben and Smeed, David and Frajka-Williams, Eleanor and Petit, Tillys and Walker, Martin and Provost, Paul and Thomas, Tina},
  journal={Geophysical Research Letters},
  volume={53},
  number={1},
  pages={e2025GL117927},
  year={2026},
  publisher={Wiley Online Library}
}

@article{johnson2024refined,
  title={Refined estimates of global ocean deep and abyssal decadal warming trends},
  author={Johnson, Gregory C and Purkey, Sarah G},
  journal={Geophysical Research Letters},
  volume={51},
  number={18},
  pages={e2024GL111229},
  year={2024},
  publisher={Wiley Online Library}
}

@article{ezer2019regional,
  title={Regional differences in sea level rise between the mid-Atlantic bight and the South Atlantic bight: is the Gulf stream to blame?},
  author={Ezer, Tal},
  journal={Earth's Future},
  volume={7},
  number={7},
  pages={771--783},
  year={2019},
  publisher={Wiley Online Library}
}

@article{wu2025gulf,
  title={Gulf Stream near Cape Hatteras modulates sea level variability along the southeastern coast of North America},
  author={Wu, Tianning and He, Ruoying},
  journal={Geophysical Research Letters},
  volume={52},
  number={7},
  pages={e2024GL112776},
  year={2025},
  publisher={Wiley Online Library}
}

@article{desbruyeres2016deep,
  title={Deep and abyssal ocean warming from 35 years of repeat hydrography},
  author={Desbruy{\`e}res, Damien G and Purkey, Sarah G and McDonagh, Elaine L and Johnson, Gregory C and King, Brian A},
  journal={Geophysical Research Letters},
  volume={43},
  number={19},
  pages={10--356},
  year={2016},
  publisher={Wiley Online Library}
}

@article{shan2022novel,
  title={A novel adaptive moving average method for signal denoising in strong noise background},
  author={Shan, Zhen and Yang, Jianhua and Sanju{\'a}n, Miguel AF and Wu, Chengjin and Liu, Houguang},
  journal={The European Physical Journal Plus},
  volume={137},
  number={1},
  pages={50},
  year={2022},
  publisher={Springer Berlin Heidelberg}
}

@article{buades2005review,
  title={A review of image denoising algorithms, with a new one},
  author={Buades, Antoni and Coll, Bartomeu and Morel, Jean-Michel},
  journal={Multiscale modeling \& simulation},
  volume={4},
  number={2},
  pages={490--530},
  year={2005},
  publisher={SIAM}
}

@article{karnauskas2021atmospheric,
  title={The atmospheric response to North Atlantic SST trends, 1870--2019},
  author={Karnauskas, Kristopher B and Zhang, Lei and Amaya, Dillon J},
  journal={Geophysical Research Letters},
  volume={48},
  number={2},
  pages={e2020GL090677},
  year={2021},
  publisher={Wiley Online Library}
}

@article{prandi2021local,
  title={Local sea level trends, accelerations and uncertainties over 1993--2019},
  author={Prandi, Pierre and Meyssignac, Benoit and Ablain, Micha{\"e}l and Spada, Giorgio and Ribes, Aur{\'e}lien and Benveniste, J{\'e}r{\^o}me},
  journal={Scientific Data},
  volume={8},
  number={1},
  pages={1},
  year={2021},
  publisher={Nature Publishing Group UK London}
}

@article{li2023recent,
  title={Recent acceleration in global ocean heat accumulation by mode and intermediate waters},
  author={Li, Zhi and England, Matthew H and Groeskamp, Sjoerd},
  journal={Nature Communications},
  volume={14},
  number={1},
  pages={6888},
  year={2023},
  publisher={Nature Publishing Group UK London}
}

@article{liao2022comparative,
  title={A comparative study of the Argo-era ocean heat content among four different types of data sets},
  author={Liao, Fanglou and Hoteit, Ibrahim},
  journal={Earth's Future},
  volume={10},
  number={9},
  pages={e2021EF002532},
  year={2022},
  publisher={Wiley Online Library}
}

@article{hogikyan2024hydrological,
  title={Hydrological cycle amplification reshapes warming-driven oxygen loss in the Atlantic Ocean},
  author={Hogikyan, Allison and Resplandy, Laure and Liu, Maofeng and Vecchi, Gabriel},
  journal={Nature climate change},
  volume={14},
  number={1},
  pages={82--90},
  year={2024},
  publisher={Nature Publishing Group UK London}
}

@article{tsujino2018jra,
  title={JRA-55 based surface dataset for driving ocean--sea-ice models (JRA55-do)},
  author={Tsujino, Hiroyuki and Urakawa, Shogo and Nakano, Hideyuki and Small, R Justin and Kim, Who M and Yeager, Stephen G and Danabasoglu, Gokhan and Suzuki, Tatsuo and Bamber, Jonathan L and Bentsen, Mats and others},
  journal={Ocean Modelling},
  volume={130},
  pages={79--139},
  year={2018},
  publisher={Elsevier}
}

@article{vancoppenolle2023si3,
  title={Si3, the nemo sea ice engine},
  author={Vancoppenolle, Martin and Rousset, C and Blockley, E and Aksenov, Y and Feltham, D and Fichefet, T and Garric, G and Gu{\'e}mas, V and Iovino, D and Keeley, S and others},
  journal={s Interneta, https://zenodo. org/record/7534900},
  volume={23},
  year={2023}
}

@article{madec2023nemo,
  title={NEMO ocean engine reference manual},
  author={Madec, Gurvan and Bell, Mike and Blaker, Adam and Bricaud, Cl{\'e}ment and Bruciaferri, Diego and Castrillo, Miguel and Calvert, Daley and Chanut, J{\'e}r{\^o}meme and Clementi, Emanuela and Coward, Andrew and others},
  journal={Mre{\v{z}}no]. Available: https://zenodo. org/record/8167700},
  volume={23},
  year={2023}
}

@article{guiavarc2025gosi9,
  title={GOSI9: UK Global Ocean and Sea Ice configurations},
  author={Guiavarc'h, Catherine and Storkey, David and Blaker, Adam T and Blockley, Ed and Megann, Alex and Hewitt, Helene and Bell, Michael J and Calvert, Daley and Copsey, Dan and Sinha, Bablu and others},
  journal={Geoscientific Model Development},
  volume={18},
  number={2},
  pages={377--403},
  year={2025},
  publisher={Copernicus Publications G{\"o}ttingen, Germany}
}

@article{bonan2025observational,
  title={Observational constraints imply limited future Atlantic meridional overturning circulation weakening},
  author={Bonan, David B and Thompson, Andrew F and Schneider, Tapio and Zanna, Laure and Armour, Kyle C and Sun, Shantong},
  journal={Nature Geoscience},
  pages={1--9},
  year={2025},
  publisher={Nature Publishing Group UK London}
}

@article{xing2026meridionally,
  title={Meridionally consistent decline in the observed western boundary contribution to the Atlantic Meridional Overturning Circulation},
  author={Xing, Qianjiang and Elipot, Shane and Johns, William E and Smeed, David A and Moat, Ben I and Loder, John W},
  journal={Science advances},
  volume={12},
  number={15},
  pages={eadz7738},
  year={2026},
  publisher={American Association for the Advancement of Science}
}

@article{ferrari2009ocean,
  title={Ocean circulation kinetic energy: Reservoirs, sources, and sinks},
  author={Ferrari, Raffaele and Wunsch, Carl},
  journal={Annual Review of Fluid Mechanics},
  volume={41},
  number={1},
  pages={253--282},
  year={2009},
  publisher={Annual Reviews}
}

@article{LineRS,
  title={Moored current and hydrographic measurements from the {RAPID WAVE Scotian Line}, 2008 to 2018},
  author={Loder,John W. and Geshelin, Yuri and Maqueda, Miguel Angel Morales and Yashayaev,Igor and Elipot, Shane and Hughes, Chris W.},
  journal={Can. Tech. Rep. Hydrogr. Ocean Sci},
  note         = {398: viii + 74 p},
  year={2025}
  }

@article{mccarthy2020sustainable,
  title={Sustainable observations of the {AMOC}: methodology and technology},
  author={McCarthy, Gerard D and Brown, Peter J and Flagg, Charles N and Goni, Gustavo and Houpert, Lo{\"\i}c and Hughes, Christopher W and Hummels, Rebecca and Inall, Mark and Jochumsen, Kerstin and Larsen, KMH and others},
  journal={Reviews of Geophysics},
  volume={58},
  number={1},
  pages={e2019RG000654},
  year={2020}
}

@article{hughes2018window,
  title={A window on the deep ocean: the special value of ocean bottom pressure for monitoring the large-scale, deep-ocean circulation},
  author={Hughes, Chris W and Williams, Joanne and Blaker, Adam and Coward, Andrew and Stepanov, Vladimir},
  journal={Progress in Oceanography},
  volume={161},
  pages={19--46},
  year={2018}
}

@article{hughes2013test,
  title={Test of a method for monitoring the geostrophic meridional overturning circulation using only boundary measurements},
  author={Hughes, Chris W and Elipot, Shane and Morales Maqueda, Miguel {\'A}ngel and Loder, John W},
  journal={Journal of Atmospheric and Oceanic Technology},
  volume={30},
  number={4},
  pages={789--809},
  year={2013}
}

@article{elipot2014observed,
  title={The observed {N}orth {A}tlantic {M}eridional {O}verturning {C}irculation: its meridional coherence and ocean bottom pressure},
  author={Elipot, Shane and Frajka-Williams, Eleanor and Hughes, Chris W and Willis, Josh K},
  journal={Journal of physical oceanography},
  volume={44},
  number={2},
  pages={517--537},
  year={2014}
}

@article{send2011observation,
  title={Observation of decadal change in the {A}tlantic meridional overturning circulation using 10 years of continuous transport data},
  author={Send, Uwe and Lankhorst, Matthias and Kanzow, Torsten},
  journal={Geophysical Research Letters},
  volume={38},
  number={24},
  year={2011}
}

@article{johns2023towards,
  title={Towards two decades of {A}tlantic {O}cean mass and heat transports at 26.5°{N}},
  author={Johns, William E and Elipot, Shane and Smeed, David A and Moat, Ben and King, Brian and Volkov, Denis L and Smith, Ryan H},
  journal={Philosophical Transactions of the Royal Society A},
  volume={381},
  number={2262},
  pages={20220188},
  year={2023},
  publisher={The Royal Society}
}

@article{lebras2023atlantic,
  title={The {A}tlantic meridional overturning circulation at 35°{N} from deep moorings, floats, and satellite altimeter},
  author={Le Bras, Isabela Alexander-Astiz and Willis, Josh and Fenty, Ian},
  journal={Geophysical Research Letters},
  volume={50},
  number={10},
  pages={e2022GL101931},
  year={2023}
}

@article{elipot2017observed,
  title={Observed basin-scale response of the {N}orth {A}tlantic meridional overturning circulation to wind stress forcing},
  author={Elipot, Shane and Frajka-Williams, Eleanor and Hughes, Chris W and Olhede, Sofia and Lankhorst, Matthias},
  journal={Journal of Climate},
  volume={30},
  number={6},
  pages={2029--2054},
  year={2017}
}

@article{elipot2013coherence,
  title={Coherence of western boundary pressure at the {RAPID} {WAVE} array: Boundary wave adjustments or deep western boundary current advection?},
  author={Elipot, Shane and Hughes, Chris and Olhede, Sofia and Toole, John},
  journal={Journal of physical oceanography},
  volume={43},
  number={4},
  pages={744--765},
  year={2013}
}

@article{bingham2008determining,
  title={Determining {N}orth {A}tlantic meridional transport variability from pressure on the western boundary: A model investigation},
  author={Bingham, Rory J and Hughes, CW},
  journal={Journal of Geophysical Research: Oceans},
  volume={113},
  number={C9},
  year={2008}
}

@article{moat2024atlantic,
  title={Atlantic meridional overturning circulation observed by the {RAPID-MOCHA-WBTS} ({RAPID}-{M}eridional {O}verturning {C}irculation and {H}eatflux {A}rray-{W}estern {B}oundary {T}ime {S}eries) array at 26°{N} from 2004 to 2023 (v2023. 1)},
  author={Moat, Ben I and Smeed, David and Rayner, Darren and Johns, William E and Smith, Ryan H and Volkov, Denis L and Elipot, Shane and Petit, Tillys and Kajtar, Jules B and Baringer, Molly O and others},
  year={2024},
  publisher={NERC EDS British Oceanographic Data Centre NOC},
  doi={doi: 10.5285/223b34a3-2dc5-c945-e063-7086abc0f274}
}

@article{toole2017moored,
  title={Moored observations of the {D}eep {W}estern {B}oundary {C}urrent in the {NW} {A}tlantic: 2004--2014},
  author={Toole, John M and Andres, Magdalena and Le Bras, Isabela A and Joyce, Terrence M and McCartney, Michael S},
  journal={Journal of Geophysical Research: Oceans},
  volume={122},
  number={9},
  pages={7488--7505},
  year={2017}
}

@article{johns2011continuous,
  title={Continuous, array-based estimates of {A}tlantic {O}cean heat transport at 26.5°{N}},
  author={Johns, William E and Baringer, Molly O and Beal, LM and Cunningham, SA and Kanzow, Torsten and Bryden, Harry L and Hirschi, JJM and Marotzke, J and Meinen, CS and Shaw, B and others},
  journal={Journal of Climate},
  volume={24},
  number={10},
  pages={2429--2449},
  year={2011}
}
\bibliographystyle{sciencemag}

%
%
%
%
%
%


\section*{Acknowledgments}
We are grateful to Prof. Chris Hughes for his assistance in extracting ocean bottom pressure along the continental slope in the ocean model, including interpreting the details of the method and providing the original IDL script.
\section*{Funding}
Q.X. was supported by U.S. National Science Foundation grant 2148723. S.E. and W.E.J. were supported by U.S. National Science Foundation grant 2148723 and 2334091. B.S., J.K., A.B., T.P., B.I.M., and D.A.S. were supported by Natural Environment Research Council grants NE/Y003551/1 and NE/Y005589/1.
\section*{Contributions}
Q.X. and S.E. designed the study. Q.X. performed the analyses, created the figures, and wrote the manuscript. S.E. and W.E.J. supervised the research and acquired funding. B.S., J.K., A.B., T.P., B.I.M., and D.A.S. contributed to data resources and interpretation. All authors reviewed and edited the paper.

\section*{Competing interests}
There are no competing interests to declare.




\newpage


\renewcommand{\thefigure}{S\arabic{figure}}
\renewcommand{\thetable}{S\arabic{table}}
\renewcommand{\theequation}{S\arabic{equation}}
\renewcommand{\thepage}{S\arabic{page}}
\setcounter{figure}{0}
\setcounter{table}{0}
\setcounter{equation}{0}
\setcounter{page}{1} 


\begin{center}
\section*{Supplementary Information for\\ \scititle}

    Qianjiang~Xing$^{1}$,
	Shane~Elipot$^{1\ast}$,
	William~E. Johns$^{1}$,
    Bablu Sinha$^{2}$, \and 
    Jules Kajtar$^{2}$,
    Adam Blaker$^{2}$,
    Tillys Petit$^{2}$,
    Ben~I. Moat$^{2}$, and
    David~A. Smeed$^{2}$
    \\
	\small$^{1}$Rosenstiel School of Marine, Atmospheric, and Earth Science, University of Miami, Miami, USA.
    \and
	\small$^{2}$National Oceanography Centre, Southampton, United Kingdom.
    \\
	\small$^\ast$Corresponding author. Email: selipot@miami.edu\and
\end{center}

\subsubsection*{This PDF file includes:}

Figures S1 to S11

\begin{figure} 
	\centering
	\includegraphics[width=1\textwidth]{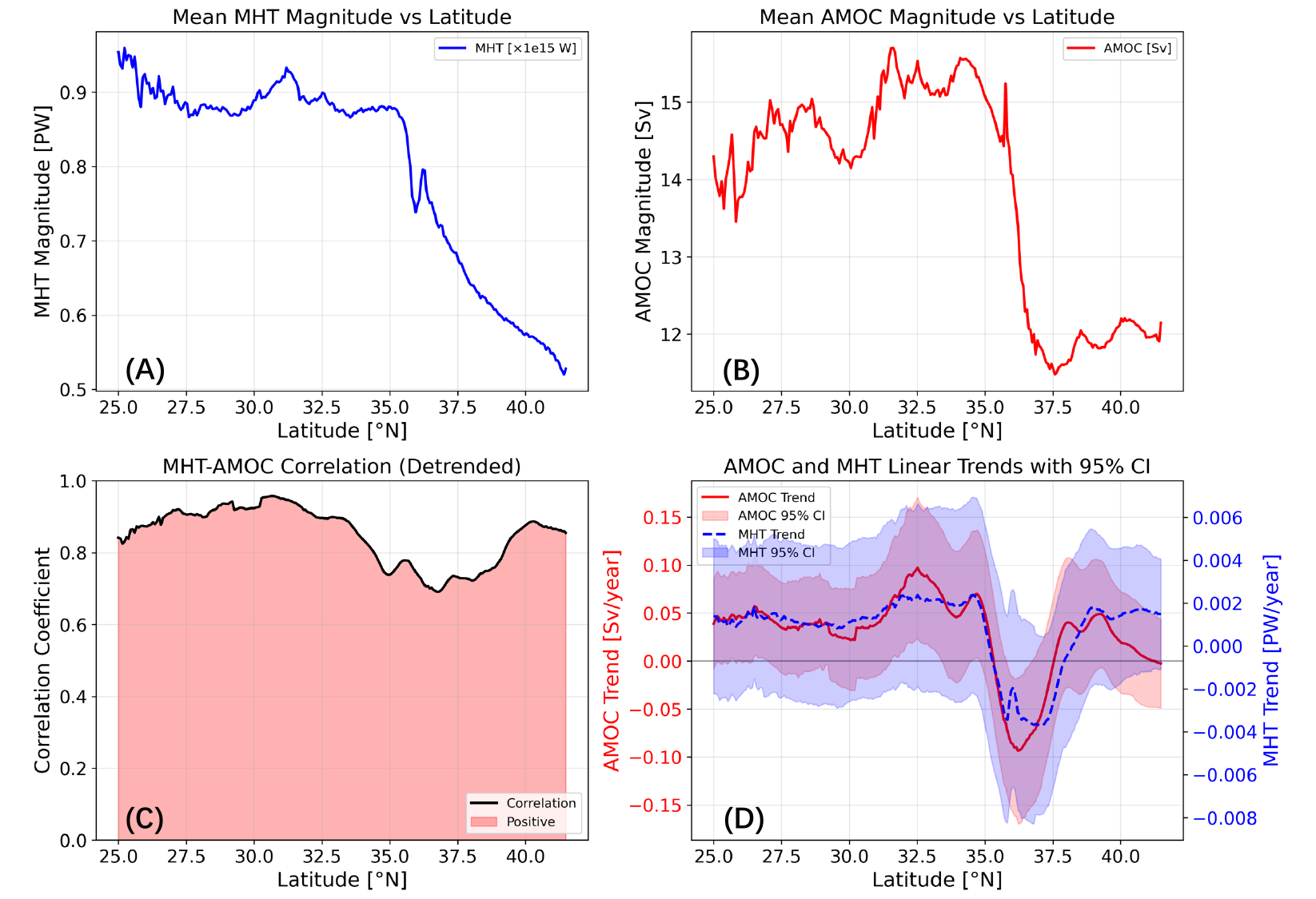} %
\end{figure}
\newpage
    \captionof{figure}{\textbf{Analysis of AMOC and MHT variability and trends in the North Atlantic (2000-2023).} . (A) Mean meridional heat transport (MHT) as a function of latitude. (B) Mean Atlantic Meridional Overturning Circulation (AMOC) strength as a function of latitude. (C) Correlation coefficient between detrended MHT and AMOC time series at each latitude. Red shading indicates positive correlation. The correlation is statistically significant at all latitudes (p $<$ 0.05). (D) Linear trends of AMOC (red, left y-axis) and MHT (blue, right y-axis) with their 95\% confidence intervals (shaded envelopes). AMOC trends are shown in solid lines with red shading, while MHT trends are shown in dashed lines with blue shading. The horizontal black line in (D) marks the zero trend line. }\label{figS1}

\begin{figure} 
	\centering
	\includegraphics[width=1\textwidth]{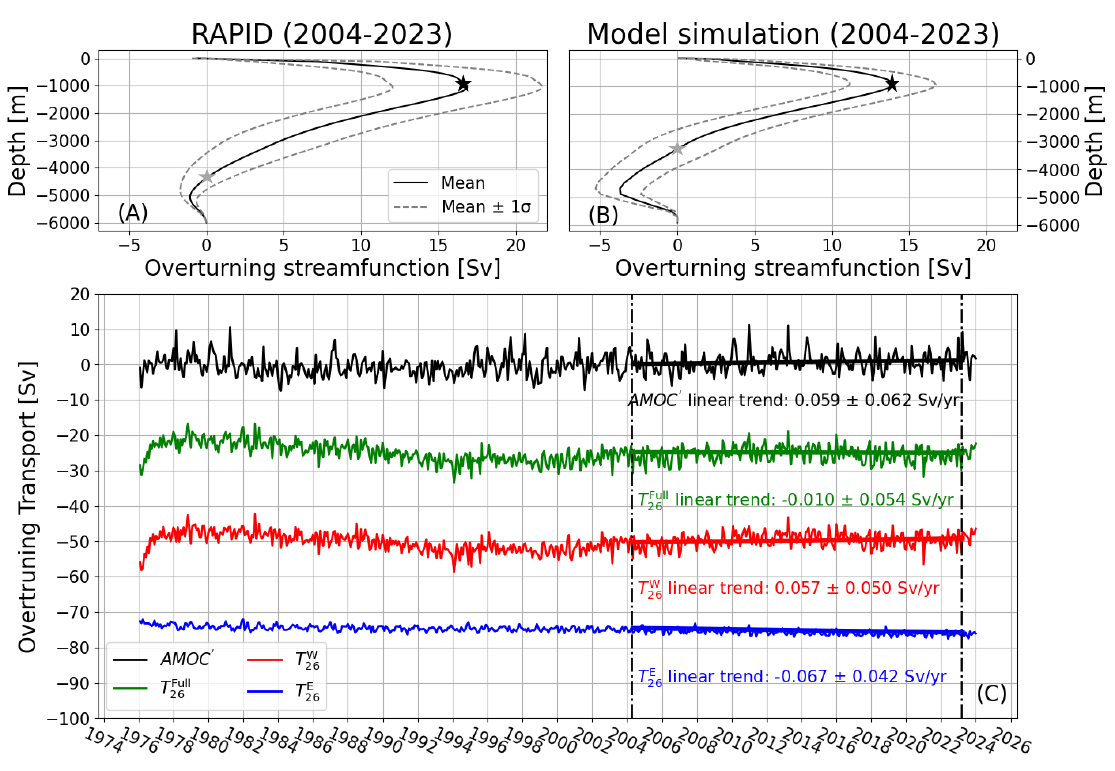} %
\end{figure}
\newpage
    \captionof{figure}{\textbf{Comparisons between AMOC estimates and deep overturning transport estimates at 26.5$^{\circ}$N.} (A) Overturning streamfunction (Sv) estimated from the RAPID array over the period 2004–2023. The solid black line represents the time-mean and the gray dashed curves indicates plus or minus one standard deviation around the time mean. Black star indicates the time-mean depth of the overturning streamfunction maximum. The time-varying maximum of the overturning streamfunction defines the Atlantic Meridional Overturning Circulation (AMOC) strength at each time step. The gray star indicates the time-mean depth of the zero-crossing of the overturning streamfunction. (B) Overturning streamfunction estimated from model simulation over the period 2004–2023. (C) Time series of simulated AMOC anomalies (black, $AMOC^{'}$), deep full overturning transport at 26.5$^{\circ}$N (green, $T_{26}^{Full}$), deep western overturning transport (red, $T_{26}^{W}$), and deep eastern overturning transport (blue, $T_{26}^{E}$) during 1976 to 2023, successively offset by -25 Sv. Note that $AMOC^{'}$ are derived from the estimated AMOC transport by changing their signs to negative for the convenience of the comparison with deep overturning transports. $T_{26}^{Full}$, $T_{26}^{W}$, and $T_{26}^{E}$ are also transport anomalies. Trend uncertainties correspond to 95\% confidence intervals. }\label{figS2}

\begin{figure} 
	\centering
	\includegraphics[width=1\textwidth]{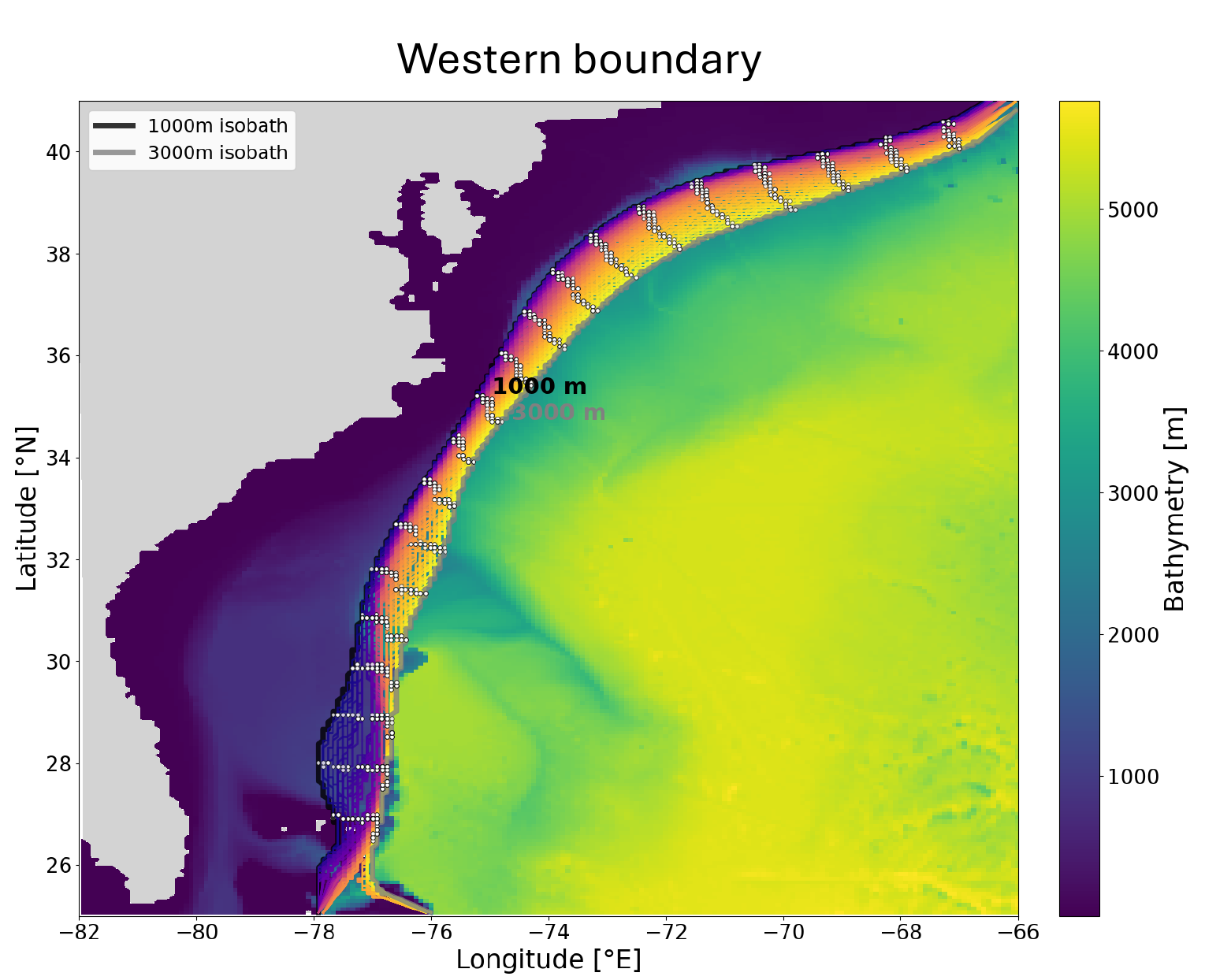} 

	\caption{\textbf{Isobath tracing and cross-slope point identification at the western boundary.} Colored contours correspond to smoothed isobaths between 1000 and 3000 m at 100 m intervals. White dots indicate the cross-slope points extracted from the smoothed isobath geometry. The line between two adjacent cross-slope points should be perpendicular to the smoothed isobaths. More details are provided in Materials and Methods.}
	\label{figS3} 
\end{figure}

\begin{figure} 
	\centering
	\includegraphics[width=1\textwidth]{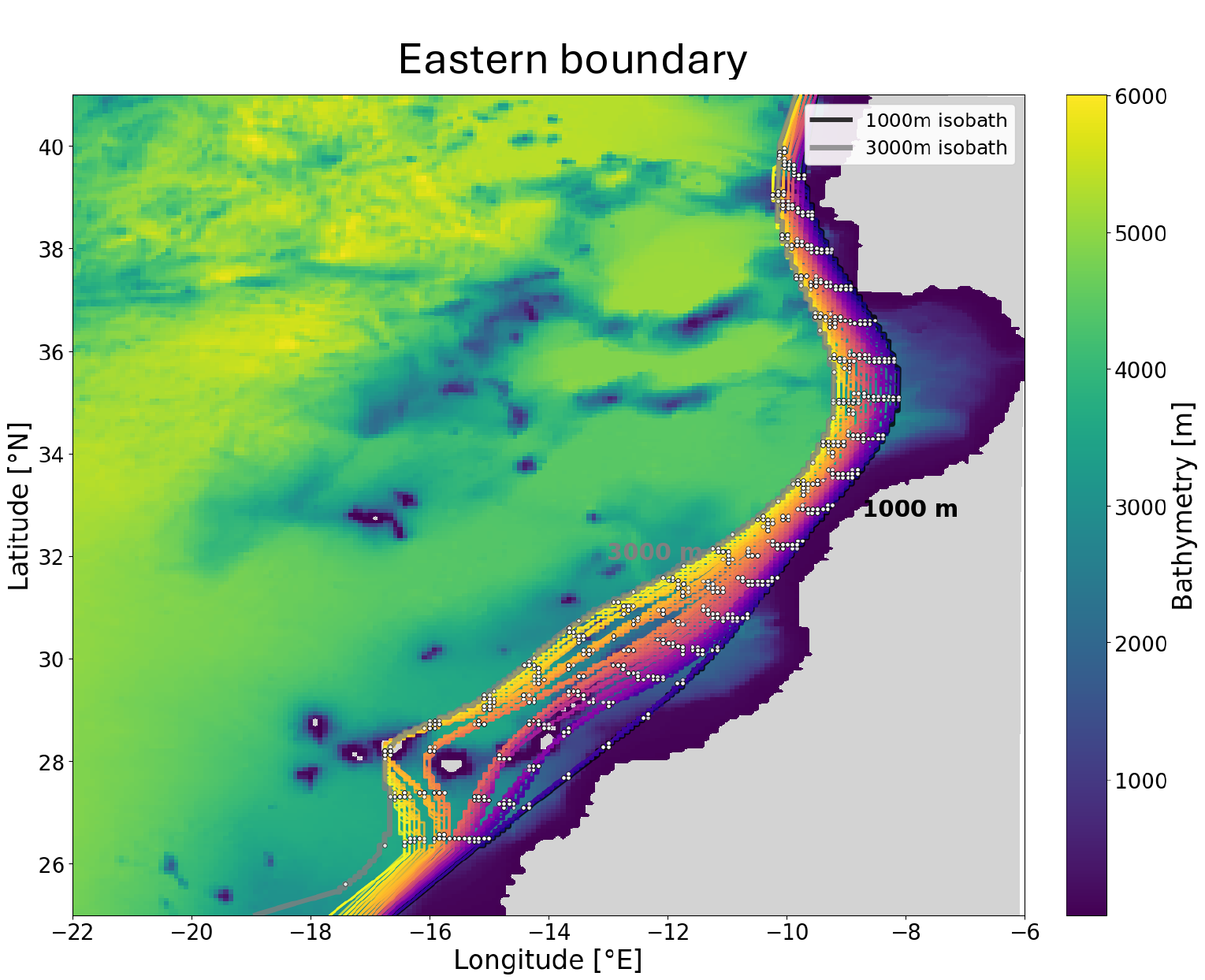} 

	\caption{\textbf{Isobath tracing and cross-slope point identification at the eastern boundary.}
		}
	\label{figS4} 
\end{figure}

\begin{figure} 
	\centering
	\includegraphics[width=1\textwidth]{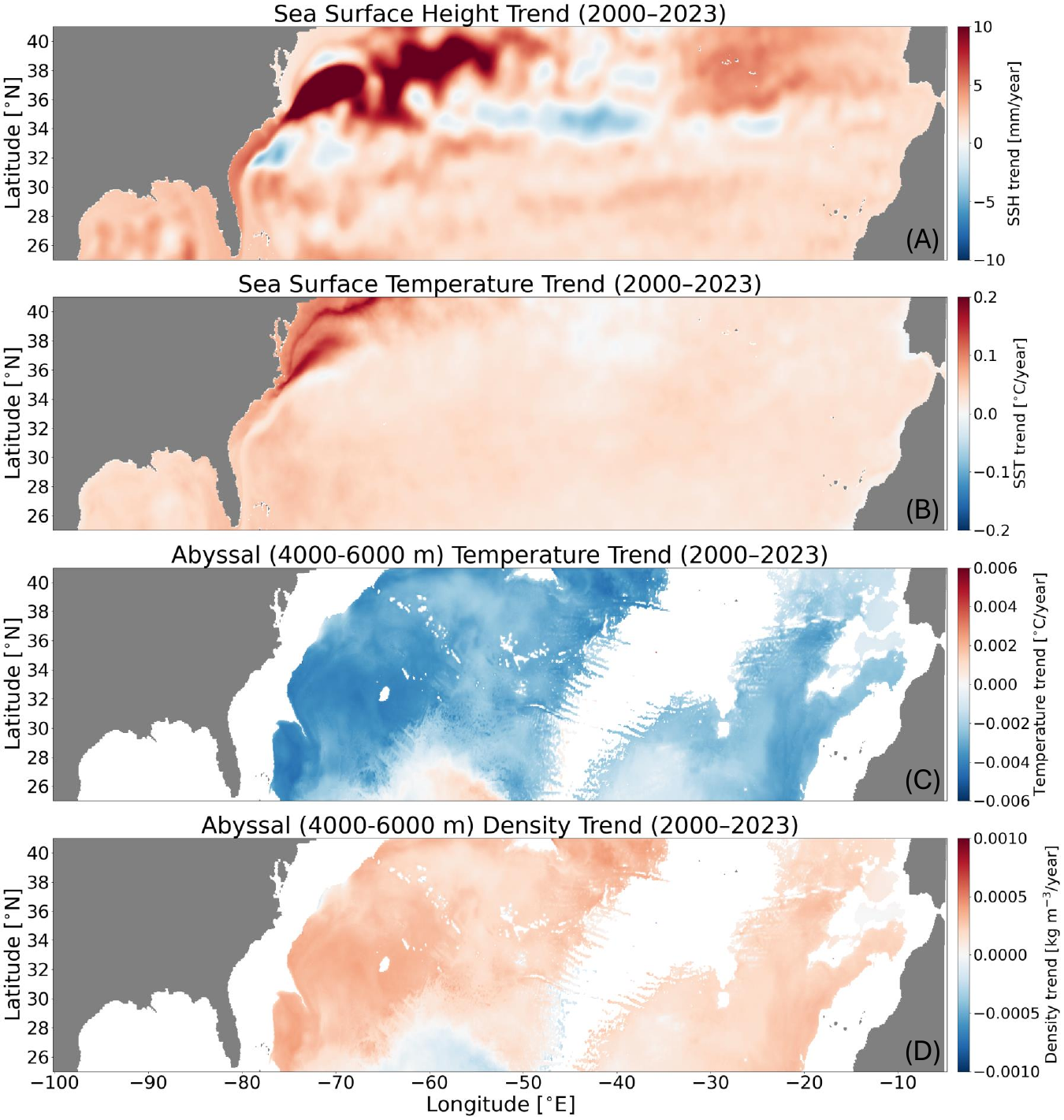} 

	\caption{\textbf{Model evaluation of climate conditions in sea surface height and sea surface temperature.}
		(A) Horizontal distribution of sea surface height trend in mid-latitude North Atlantic (25-41$^{\circ}$N) during 2000-2023. (B) Horizontal distribution of sea surface temperature trend during 2000-2023. (C) and(D) Horizontal distribution of abyssal (4000-6000 m) temperature and density trend during 2000-2023.}
	\label{figS5} 
\end{figure}

\begin{figure} 
	\centering
	\includegraphics[width=1\textwidth]{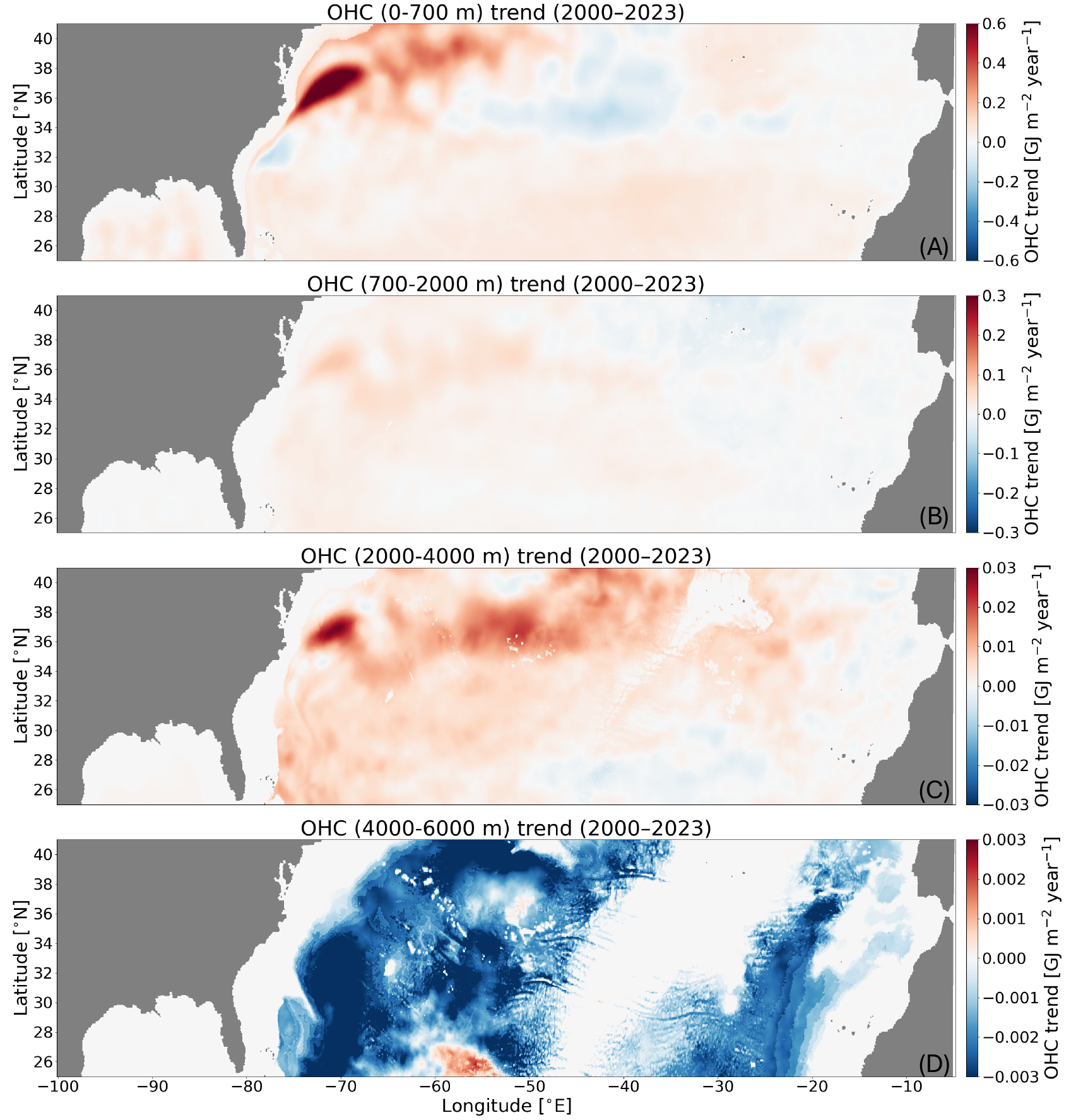} 

	\caption{\textbf{Model evaluation of climate conditions in ocean heat content.}
		(A) Horizontal distribution of the vertically integrated ocean heat content (OHC) trend in the upper 700 m of the mid-latitude North Atlantic during 2000–2023. (B) Horizontal distribution of the vertically integrated OHC trend for 700–2000 m. (C) Horizontal distribution of the vertically integrated OHC trend for 2000–4000 m. (D) Horizontal distribution of the vertically integrated OHC trend below 4000 m. Note that the colorbar scales are not uniform across panels and reflect the magnitude of OHC trends in each depth range.}
	\label{figS6} 
\end{figure}

\begin{figure} 
	\centering
	\includegraphics[width=1\textwidth]{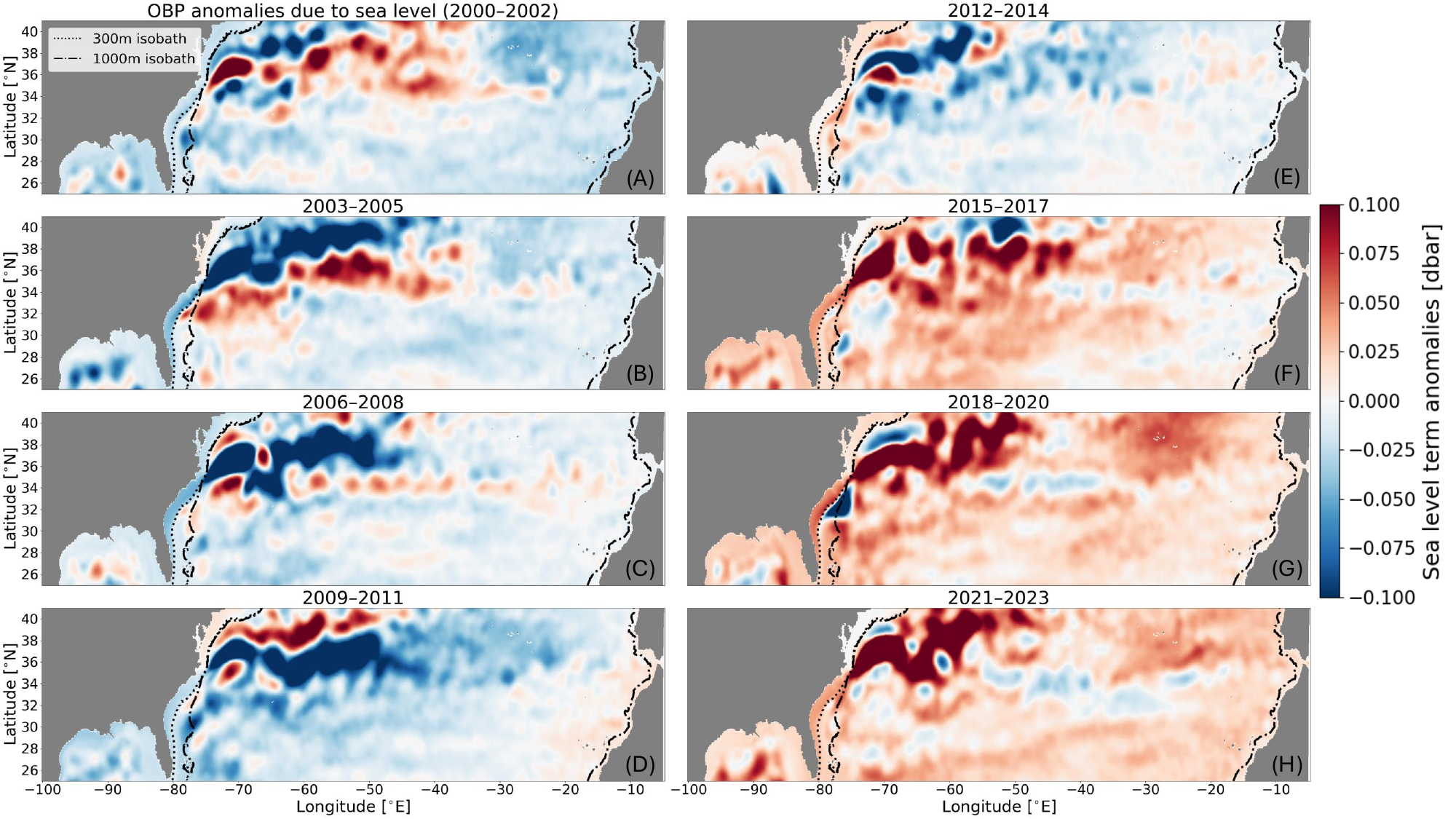} 

	\caption{\textbf{Evolution of OBP anomalies due to sea level in the mid-latitude North Atlantic during 2000–2023.} (A-H) OBP anomalies respect to the time mean due to sea level change in consecutive 3-year intervals from 2000 to 2023.}
	\label{figS7} 
\end{figure}
\clearpage 

\begin{figure} 
	\centering
	\includegraphics[width=1\textwidth]{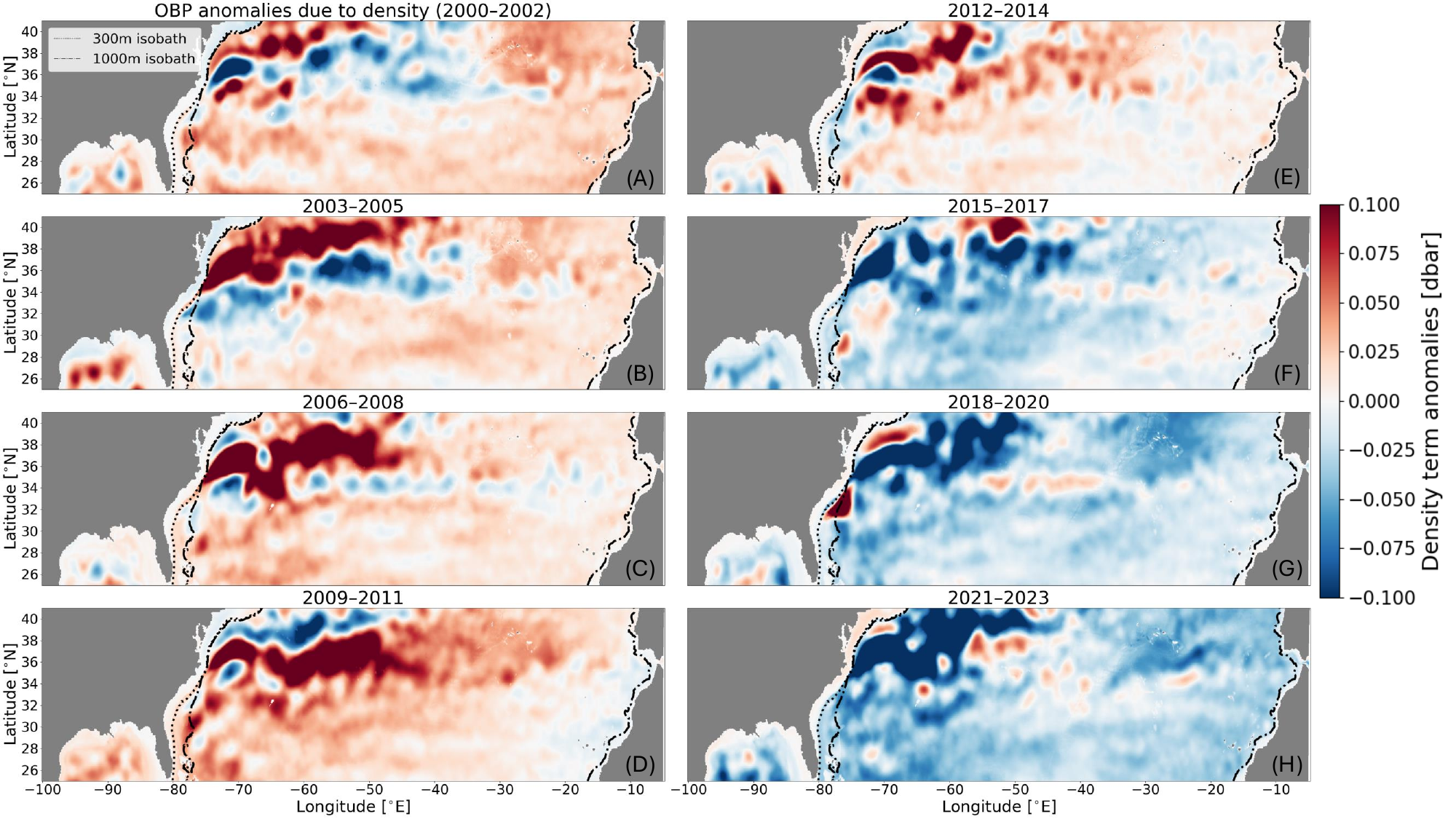} 

	\caption{\textbf{Evolution of OBP anomalies due to density in the mid-latitude North Atlantic during 2000–2023.} (A-H) OBP anomalies with respect to the time mean due to density change in consecutive 3-year intervals from 2000 to 2023.}
	\label{figS8} 
\end{figure}

\begin{figure} 
	\centering
	\includegraphics[width=1\textwidth]{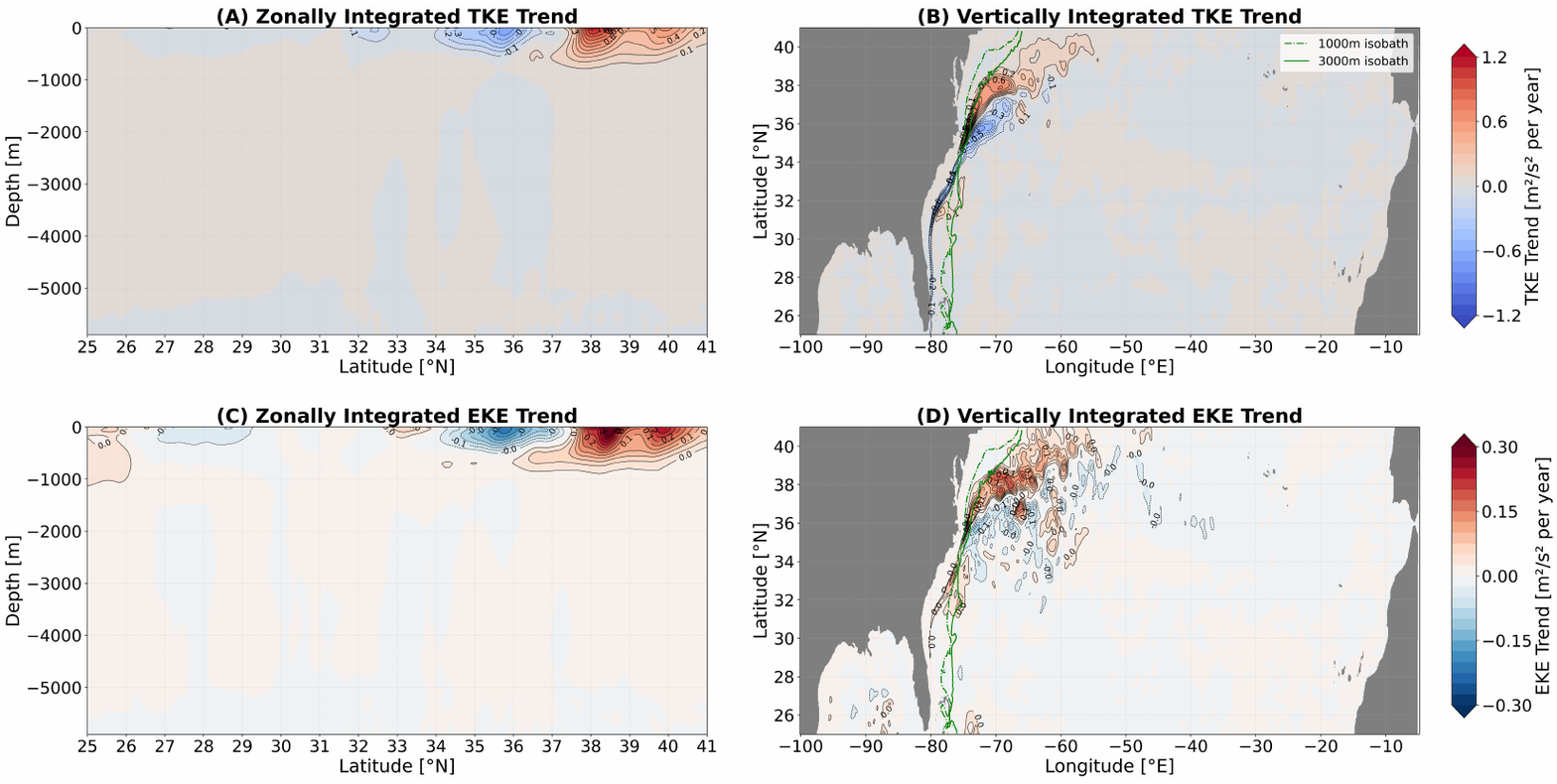} 

	\caption{\textbf{Trends of total kinetic energy (TKE) and eddy kinetic energy (EKE) in the North Atlantic (2000–2023).} (A) Zonally integrated TKE trend as a function of latitude and depth. (B) Vertically integrated TKE trend over the mid-latitude North Atlantic. Green lines show the 1000 and 3000 m isobath along the western boundary. (C) Same as (A) but for EKE trend. (D) Same as (b) but for EKE trend. The positive EKE trend near Cape Hatteras reflects a strengthening of localized eddy-induced transport, contributing to a regional intensification of the residual circulation.}
	\label{figS9} 
\end{figure}
\clearpage

\begin{figure} 
	\centering
	\includegraphics[width=1\textwidth]{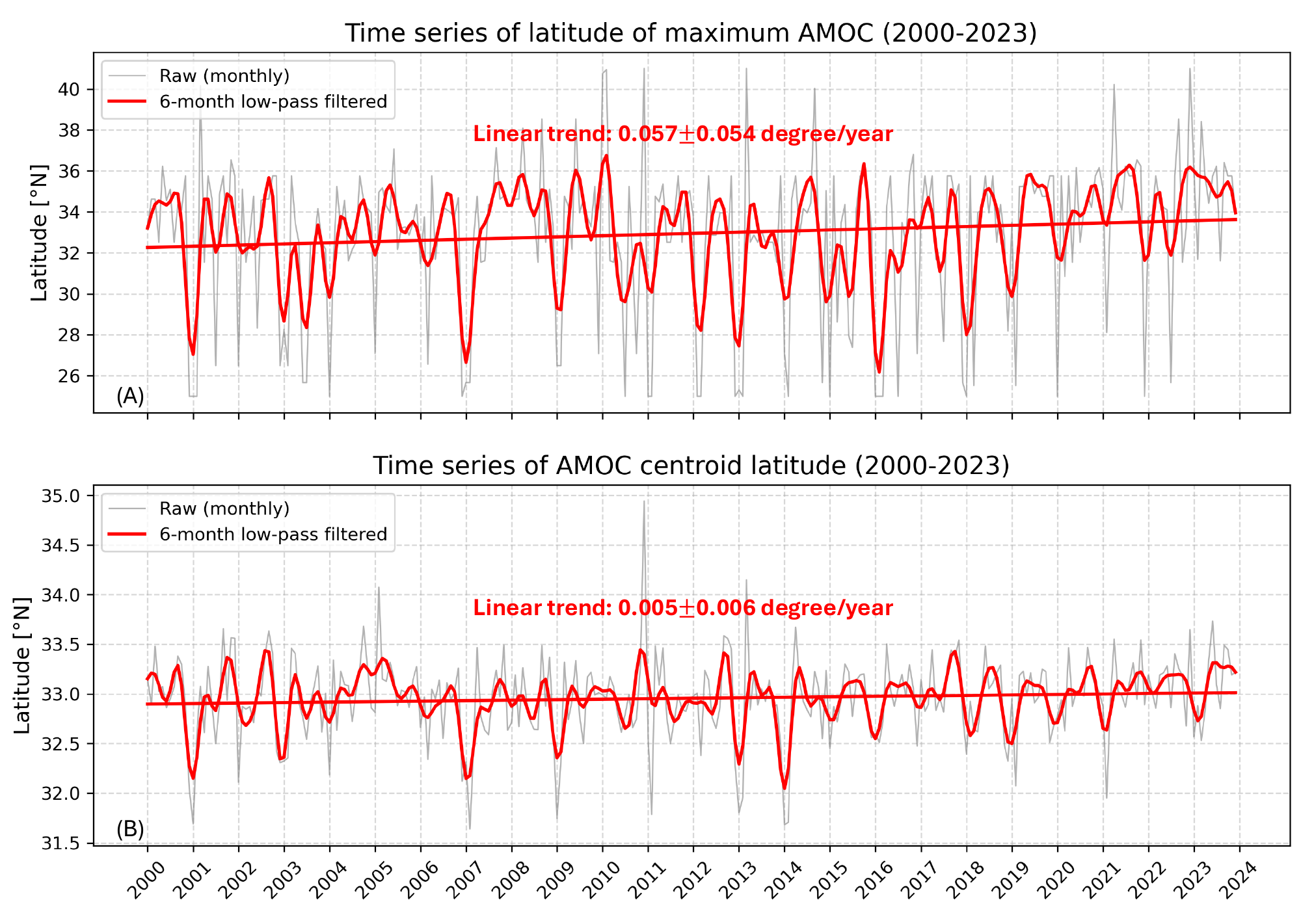} 

	\caption{\textbf{Northward shift of AMOC during 2000-2023.} (A) Time series of the latitude of maximum AMOC (at approximately 1000 m depth) over the period 2000–2023. (B) Time series of the AMOC centroid latitude over the period 2000–2023. The gray line shows the monthly time series, and the red line represents a 3-month, third-order Butterworth low-pass filtered version. The maximum AMOC is identified as the maximum value of the AMOC streamfunction across all latitudes of the entire basin. The AMOC centroid latitude is defined as the latitude-weighted average of the AMOC streamfunction. The linear trend is calculated based on the filtered version, and the corresponding uncertainties represent 95\% confidence intervals.}
	\label{figS10} 
\end{figure}
\clearpage

\begin{figure} 
	\centering
	\includegraphics[width=1\textwidth]{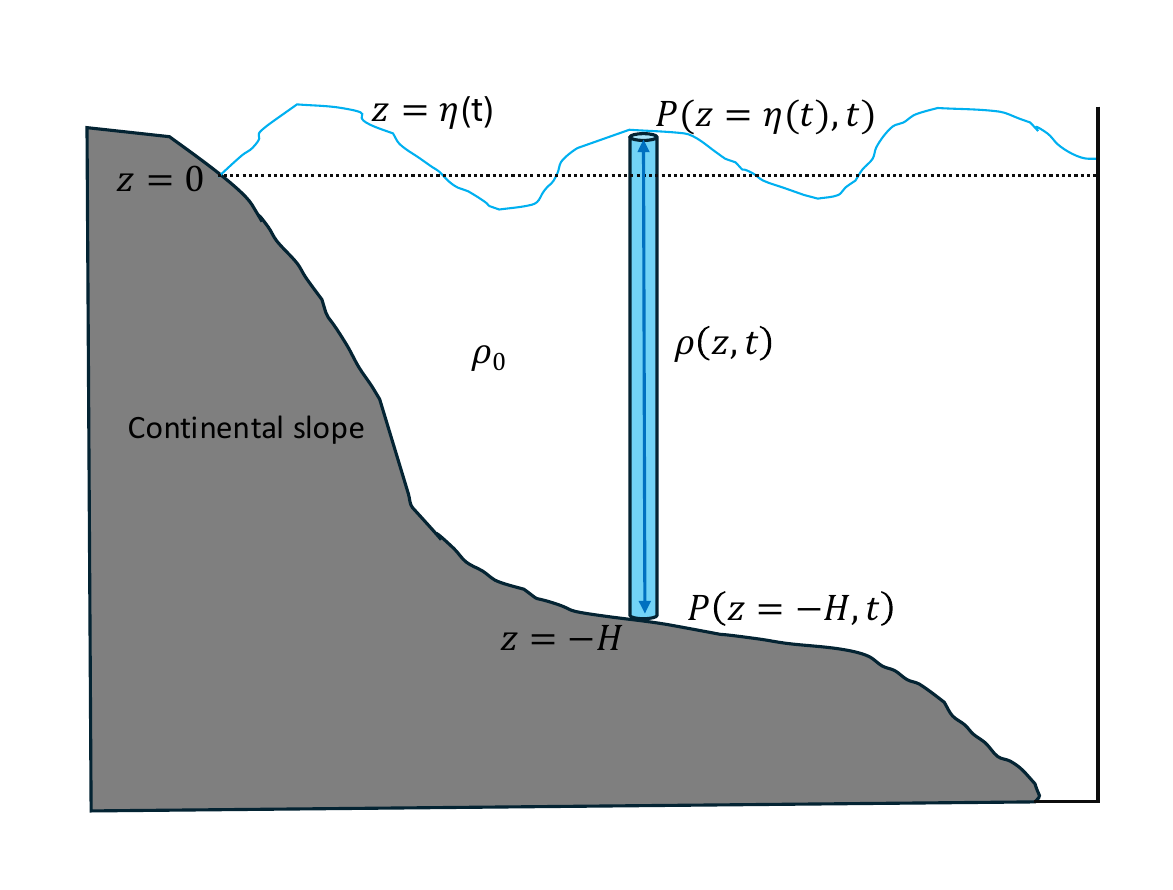} 

	\caption{\textbf{Schematic illustrating the decomposition of the full water column on the continental slope.} $P(z = -H, t)$ is OBP on the continental slope, at time $t$ and depth $z=-H$,   $\rho(z,t)$ is the ocean water density, $g$ is the gravitational acceleration, $\eta(t)$ is the time-varying sea surface height relative to the geoid ($z=0$). $P(z=\eta(t), t)$ is the surface pressure, and $\rho_0$ is temporally and spatially constant ocean water density.}
	\label{figS11} 
\end{figure}
\clearpage



\end{document}